\documentclass[twocolumn]{aastex631}
\usepackage{graphicx}
\accepted{for publication in AJ}
\shorttitle{3D orbital architecture of a dwarf binary system}
\shortauthors{Curiel et al.}

\begin{document} 

%
%
\title{3D orbital architecture of a dwarf binary system and its planetary companion}

\correspondingauthor{Salvador Curiel}
\email{scuriel@astro.unam.mx}
\author[0000-0003-4576-0436]{Salvador Curiel$^{*}$}
\affil{Instituto de Astronom{\'\i}a,
Universidad Nacional Aut\'onoma de M\'exico (UNAM),
Apdo Postal 70-264,
Ciudad de M\'exico, M\'exico.
scuriel@astro.unam.mx}

\author[0000-0002-2863-676X]{Gisela N. Ortiz-Le\'on$^{1}$}
\affil{Max Planck Institut f\"ur Radioastronomie,
Auf dem H\"ugel 69, D-53121 Bonn, Germany}


\author[0000-0002-2564-3104]{Amy J. Mioduszewski}
\affiliation{National Radio Astronomy Observatory,
P.O. Box 0, Socorro, NM 87801, USA }

\author[0000-0002-9723-0421]{Joel Sanchez-Bermudez$^{1}$}

\begin{abstract}

\noindent
Because of the diversity of stellar masses and orbital sizes of binary systems, and the complex interaction between star-star, star-planet and planet-planet, it has been difficult to fully characterize the planetary systems associated with binary systems.
Here, we report high-precision astrometric observations of the low-mass binary system GJ~896AB, revealing the presence of a Jupiter-like planetary companion (GJ~896A$b$).
The planetary companion is associated to the main star GJ~896A, with an estimated mass of 2.3 Jupiter masses and an orbit period of 284.4 days.
A simultaneous analysis of the relative astrometric data obtained in the optical and infrared with several telescopes, and the absolute astrometric data obtained at radio wavelengths with the Very Long Baseline Array (VLBA), reveals, for the first time, the fully characterized three-dimensional (3D) orbital plane orientation of the binary system and the planetary companion. The planetary and binary orbits are found to be in a retrograde configuration and with a large mutual inclination angle ($\Phi$ = 148$^\circ$) between both orbital planes. 
Characterizing the 3D orbital architecture of binary systems with planets is important in the context of planet formation, as it could reveal whether the systems were formed by disk fragmentation or turbulence fragmentation, as well as the origin of spin-orbit misalignment.
Furthermore, since most stars are in binary or multiple systems, our understanding of systems such as this one will help to further understand the phenomenon of planetary formation in general. 

\end{abstract}




\section{Introduction} \label{sec:intro}

Numerical simulations suggest two main channels for the formation of multiple stellar systems: disk fragmentation \citep[][]{adams89}, which produces secondaries through gravitational instability within a massive accretion disk, and turbulent fragmentation \citep[e.g.,][]{goodwin04, fisher04}, where turbulence in the original molecular core leads to multiple density enhancements, which independently collapse. 
Binary systems formed by turbulence fragmentation are expected to have initial separations larger than 500 astronomical units (au), which corresponds to the rough boundary between the disk size and the molecular core scale \citep[e.g.,][]{offner10}, while disk fragmentation suggest the formation of compact ($<$ 500 au) binaries  \citep[e.g.,][]{kratter10}.
However, dynamical evolution may quickly modify the separation, making that initially wide binaries migrate to separations $<$200 au in time spans of $\sim$0.1 Myr \citep[][]{offner10, offner16}. If substantial orbital evolution occurs during the main accretion phase, which lasts $\sim$0.5 Myr \citep[][]{dunham14}, then binary systems will end up much closer binary systems.

Theoretical works have addressed the formation of planets around single stars of different masses. 
The core-accretion theory predicts that the formation of giant-mass planets scales with the mass of the central star; thus, it is expected that very few Jovian-mass planets are formed around low-mass stars \citep[e.g.,][]{laughlin04,kennedy08,mercer20}.
The core-accretion theory indicates that these planets would be formed in orbits far from the star, at several au. On the other hand, it is expected that disk fragmentation may also be able to form giant-mass planets around low-mass stars  \citep[e.g.,][]{boss06}. In this case, the orbit of the planet is expected to be relatively closer to the star, from a few to several au.

Only a very small fraction of the detected extrasolar planets (less than 4\%) are known to be associated to stars in binary systems. This is probably, at least in part, due to strong observational biases. For instance, the radial velocity (RV) technique is in general limited to binary systems with separations larger than 2$''$, and the transit technique is limited to high inclination angles of the orbital plane of the planets and the binary systems \citep[e.g.,][and references therein]{marzari19}. In addition, the presence of a stellar companion has adverse effects on planet formation. It is possible that the stellar companion strongly influence both the formation of planets and their subsequent dynamical evolution. For example, the presence of a close stellar companion can tidally truncate the protoplanetary disk \citep[e.g.,][]{savonije94, kraus12}. Observational surveys have shown that, even when present, protoplanetary disks are on average less massive in tight binaries \citep[e.g.,][]{harris12}. In addition, the evolution of truncated disks under the action of viscous forces is faster and should produce more short-lived disks, with less time available for planet formation \citep[e.g.,][]{muller12}. This could particularly affect the formation of giant gaseous planets, which need to accrete vast amount of gas before the primordial disk disperses. There is observational evidence that suggests that the typical life time of protoplanetary disks in close binaries separated by $\leq$40 au is less than 1 Myr, as compared with the typical disk life time of 5$-$10 Myrs in single stars \citep[e.g.,][]{marzari19}. Furthermore, in the case of close binaries, there might not be enough mass left in the truncated disks to form jovian planets. Only a small fraction (about 15\%) of the planets associated with binaries have been found associated to binary systems with separations $\leq$40 au \citep[e.g.,][]{marzari19, fontanive21}, and according to the Catalogue of Exoplanets in Binary Systems \citep[][]{schwarz16}
only 7 planets are associated to M dwarf binary systems. The formation of planets in binary systems is not well understood. Current planetary formation models take into account only the formation of planets around single stars. To our knowledge, there are no theoretical models that take into account, for instance, the simultaneous formation of planetary and sub-stellar or stellar companions in a single protoplanetary disk, or the formation of planetary systems during the formation of a binary system through turbulent fragmentation. 


A variety of mechanisms has been proposed to explain spin-orbit misalignment observed in several planetary systems. Dynamical interactions between stars and/or planets \citep[][]{nagasawa08}  help explaining small and large differences in the inclination angle of close-in giant-planets. The chaotic star formation environment during the accretion phase \citep[][]{bate10} and perturbations from a stellar companion \citep[][]{thies11, batygin13, lai14} also help explaining large misalignment between the stellar spin and the planetary orbit. 
Thus, studying the spins of individual members in binary systems (including sub-stellar and planetary companions) could reveal whether they were formed by disk fragmentation or turbulence fragmentation, as well as the origin of the spin-orbit misalignment \citep[e.g.,][]{offner16}. 
To accomplish this, the three-dimensional (3D) orbital plane orientation of both, the binary system and the planetary companions would need to be known. 
However, in most cases, not all angles of the orbital plane of the binary system and their planetary companions are known. In all known binary systems with planetary companions, the position angle of the line of nodes ($\Omega$) is unknown, and in several cases even the inclination angle of the orbits is also unknown.
Binaries formed within the same accretion disk are likely to have common angular momenta and, therefore, aligned stellar spins, whereas binaries formed via turbulent fragmentation are likely to possess independent angular momentum vectors and, thus, have randomly oriented spins. It is expected that the planets have orbital spins similar to the stellar spin of their host star. However, in the case of binary systems, the orientation and eccentricity of the planetary orbits are expected to be affected by the presence of the stellar companion. 

Astrometry is the only technique capable of directly giving all three angles (longitude of the periastron $\omega$, position angle of the line of nodes $\Omega$, and the inclination angle $i$) of the orbital planes. In particular, radio interferometric astrometry, with milli-arc-second (mas) angular resolution and very high astrometric precision (usually better than 60 micro-arc-second ($\mu$as)), can be used to search for planetary companions associated to binary systems with wide orbits, as well as close binary systems with separations $\leq$50 au.

\subsection{GJ~896 binary system} \label{sec:bsys}

GJ~896AB (a.k.a. EQ Peg, BD+19~5116, J23318+199, HIP 116132) is a nearby low-mass M dwarf binary system with an estimated age of 950 Myr \citep[][]{parsamyan95}, at a distance of 6.25 pc,  and a separation of 5.4 arcsec with a PA of 78$^\circ$ \citep[e.g.,][]{heintz84, liefke08, bower11, pearce20}. However, some observations also suggest that the age of this binary system is $\lesssim$ 100 Myr \citep[][]{riedel11, zuckerman13}. If this latter age is confirmed, these stars would be the closest pre-main-sequence stars known. This binary system is part of a quadruple system, where the other two stellar companions are further away from the main binary. Recent observations suggest that both stars have companions, but theirs orbits have not been characterized  \citep[e.g.,][]{delfosse99, winters21}.

Both stars have been observed to have radio outbursts \citep[e.g.,][]{pallavicini85, jackson89, benz95, gagne98, crosley18a, crosley18b, villadsen19, davis20} and are flaring X-ray emitters \citep[e.g.,][]{robrade04, liefke08}. 
GJ~896A has been detected at several epochs with the VLBA showing compact and variable flux emission \citep[][]{bower09, bower11}. 

An early determination of the orbital motion of the binary GJ~896AB suggested an orbital period of about 359 yrs and a semi-major axis of 6.87 arcsec \citep[e.g.,][]{heintz84}.
More recent, and more accurate, relative astrometric fits indicate that the orbital parameters of this binary system are $P$ $\sim$234 yr, $a$ $\sim$ 5.3 arcsec, $i$ $\sim$ 126$^\circ$, $\Omega$ $\sim$ 77$^\circ$, $e$ $\sim$ 0.11 and  $\omega$ $\sim$ 97$^\circ$ \citep[][]{mason01}. However, since the time baseline used for these orbital determinations is $\lesssim$34\% the estimated period, the orbital parameters were not well constrained.
The more massive star, GJ~896A, is an M3.5 star with an estimated mass of 0.39 M$_\odot$, a radius of 0.35 R$_\odot$, a rotation period of 1.061 days, and a rotational inclination angle of 60$^{\circ}$, while the less massive star, GJ~896B, is an M4.5 star with an estimated mass of 0.25 M$_\odot$, a radius of 0.25 R$_\odot$, a rotation period of 0.404 days, and a rotational inclination angle of about 60$^{\circ}$$\pm$20$^{\circ}$ \citep[e.g.,][]{delfosse99, morin08, davison15, pearce20}.

Here we present the discovery of a Jovian-mass planetary companion to the young close-by (6.25 pc away from the Sun) M3.5 dwarf GJ~896A, which is the more massive star in the low-mass M dwarf binary system GJ~896AB with a separation of about 31.6 au. 
Section~\ref{sec:obs} describes the observations and data reduction. In Section~\ref{sec:procid}  we present the methodology to fit the astrometric data. Section~\ref{sec:results} presents the results, including the fitted 3D orbital architecture of this system. The results are discussed in Section~\ref{sec:discusion}  and our conclusions are given in Section~\ref{sec:conclusions}.
In Appendix~\ref{sec:flares} we discuss the possible contributions of the variability of the main star to the expected $``$jitter$"$ of the star, and in Appendix~\ref{sec:mcmc} we present the posterior sampling of the combined astrometric fit solution using an MCMC sampler.

\section{Observations and Data Reduction} \label{sec:obs}

We analyzed archival and new Very Long Baseline Array (VLBA) observations taken at 8.4~GHz toward the M dwarf binary system GJ~896AB. Observations of GJ~896A  were carried out in fourteen epochs between March, 2006 and November, 2011 as part of the Radio Interferometric Planet  (RIPL) survey and its precursor (\citealt{bower09,bower11}; program IDs: BB222 and BB240). New observations of the two stars GJ~896A and GJ~896B were obtained as part of our own program BC264 (PI: S.\ Curiel) in three epochs between August, 2020 and October, 2020. The RIPL observations recorded  four  16-MHz frequency bands in dual-polarization mode. For the most recent observations, we recorded  four 128-MHz frequency bands, also in dual polarization  mode, using the new  4~Gbps recording rate of the VLBA. The observing sessions consisted of switching scans between the target and the phase reference calibrator, J2328+1929, spending about 1~min on the calibrator and 3~min on the target.  Scans on secondary calibrators (J2334+2010, J2334+1843 and J2328+1956) were obtained every $\approx$30--60 min. In addition, fringe calibrators were observed occasionally during the sessions.  For program  BC264, additional 30-min blocks of calibrators, the so-called geodetic-like blocks,  distributed over a wide range of elevations were included at the beginning and end of the observing run. The RIPL observations included the 100-m Green Bank Telescope added to the VLBA array. 
In project BC264 the typical on source time is of 2 hours, spread over a full track of 3 hours, plus 1 hour for the two geodetic-like blocks. On the other hand, RIPL alternated between two targets in a 8 hours track, and no geodetic-like blocks were observed.

To determine the orbital motion of the binary system, we used archival relative astrometric measurements of GJ~896AB taken from the  Washington Double Star Catalog \citep{mason01}, maintained by the US Naval Observatory.  This data set includes a total of 73 measurements starting in the year 1941 until 2017. We discarded the data points from 1950 and 1952.66 since they largely deviate from the rest of the observations due to possible systematics. 

The VLBA data were reduced with the Astronomical Imaging System \citep[AIPS;][]{greisen03}, following standard procedures for phase-referencing observations \citep{torres07,ortizleon17}. 
First, corrections for the ionosphere dispersive delays were applied. We then corrected for post-correlation updates of the Earth orientation parameters. Corrections for the digital sampling effects of the correlator were also applied. The instrumental single-band delays caused by the VLBA electronics, as well as the bandpass shape corrections were determined from a single scan on a strong fringe calibrator and then applied to the data. Amplitude calibration was performed by using the gain curves and system temperature tables to derive the system equivalent flux density of each antenna. We then applied corrections to the phases for antenna parallactic angle effects. Multi-band delay solutions were obtained from the geodetic-like blocks, which were then applied to the data to correct for tropospheric and clock errors.  The final step consisted of removing global frequency- and time-dependent residual phase errors obtained by fringe-fitting the phase calibrator data, assuming a point source model. In order to take into account the non-point-like structure of the calibrator, this final step was repeated using a self-calibrated image of  the calibrator as a source model. Finally, the calibration tables were applied to the data and images of the stars were produced using the CLEAN algorithm. We used a pixel size of 50~$\mu$as and pure natural weighting. The synthesized beam of  these images are, on average, $2.7\times 1.1$~mas, and the achieved rms noise levels are in the range between 11 and 130~$\mu$Jy~beam$^{-1}$ (Table \ref{tab_1}). The large range of sensitivities is expected because of the wide range in observing strategies.

The assumed position for the primary phase calibrator J2328+1929 during correlation changed between epochs due to multiple updates of the VLBA calibrator position catalogs. Thus, before deriving any calibration, the position of J2328+1929 was corrected to the value assumed in the observations of 2020, R.A.$=$23:28:24.874773 and Dec.$=$+19:29:58.03010. The assumed positions for the observations performed between 2006 and 2011 were taken from the correlator files available in the VLBA file
server at {http://www.vlba.nrao.edu/astro/VOBS/astronomy/}.

\begin{figure*}
\begin{center}
  \plotone{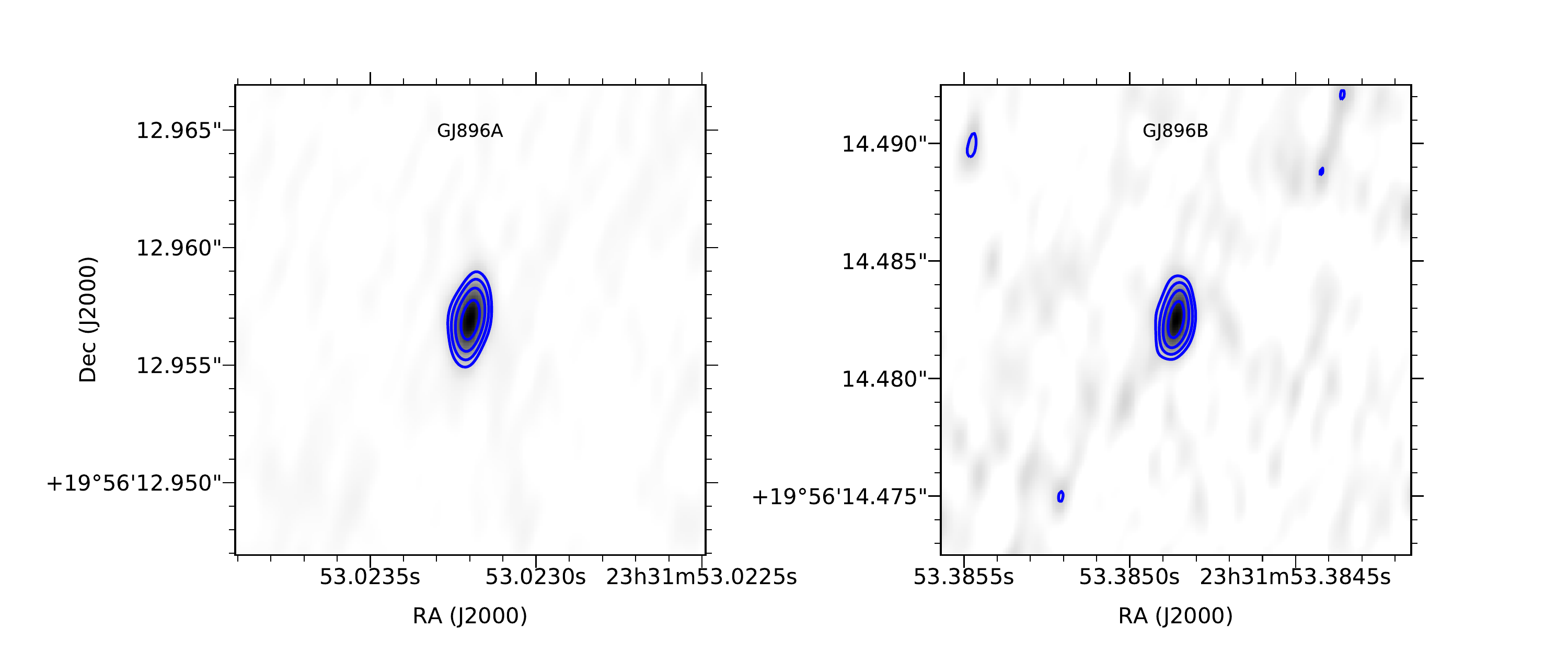}
\end{center}
  \caption{The intensity maps of GJ~896A (left panel) and GJ~896B (right panel) taken on September 13, 2020 are shown here as an example. 
  The $n$th  contour is at $\left({\sqrt{2}}\right)^{n}\times S_{\rm max} \times p$, where $S_{\rm max}$ is the peak flux in the image (0.57 mJy for GJ~896A and 0.16 mJy for GJ~896B, respectively), $n$=1, 2 ..., and $p$ is equal to 20\%.}
  \label{fig_1}%
\end{figure*}

GJ~896A was detected in thirteen epochs of RIPL and in the three epochs of program BC264. GJ~896B was only detected in two epochs of BC264. 
To obtain the positions of the centroid in the images of GJ~896A and B, we used the task {\tt MAXFIT} within AIPS, which gives the position of  the pixel with the peak flux density. The position error is given by the astrometric uncertainty, $\theta_{\rm res}/(2\times {\rm S}/{\rm N})$, where $\theta_{\rm res}$ is the full width at half maximum (FWHM) size of the  synthesized beam, and ${\rm S}/{\rm N}$ the signal-to-noise ratio of the source \citep{thompson17}. Furthermore, we quadratically added half of the pixel size to the position error. 
In order to investigate the magnitude of systematic errors in our data, we obtain the positions of the secondary calibrator, J2334+2010, in all observed epochs. The rms variation of the secondary calibrator position is (0.14, 0.14)~mas. The angular separation of J2334+2010 relative to the main calibrator is $1\rlap.{^{\rm o}}6$, while the target to main calibrator separation is $0\rlap.{^{\rm o}}97$. The main calibrator, target and secondary calibrator are located in a nearly linear arrangement (see Fig.\ 1 in \citealt{bower11}). Since systematic errors in VLBI phase-referenced observations scale linearly with the source-calibrator separation \citep{pradel06,reid14}, we scale the derived   rms value for J2334+2010 with the ratio of the  separations from the target and J2334+2010 to the main calibrator. This yields a systematic error of (0.09, 0.08)~mas, which was added in quadrature to the position errors in each coordinate. 
Table \ref{tab_1} summarizes the observed epochs, the positions, the associated uncertainties, and the integrated flux densities of GJ~896A and B. 
Figure~\ref{fig_1} shows the intensity maps of both stars for one of the two epochs when both stars were detected with the VLBA.
The time of the observations (in Julian day)  included in Table \ref{tab_1} corresponds to the average time of the time span of each observed epoch (typically of about 4 hours, including the geodetic-like blocks in BC264, and 8 hours in RIPL). The integrated  flux densities were obtained by fitting the source brightness distribution with a Gaussian model.

\section{Fitting of the Astrometric Data.}  \label{sec:procid}

We followed the same fitting procedure presented by \citet[][]{curiel19, curiel20}. We used two astrometric fitting methods: non-linear Least-squares algorithm \citep[][]{curiel19, curiel20} and the asexual genetic algorithm AGA \citep[][]{canto09, curiel11, curiel19, curiel20}. 
Both fitting codes include iterative procedures that search for the best fitted solution in a wide range of posible values in the multi-dimensional space of parameters. These iterative procedures help the fitting codes not be trapped in a local minimum, and to find the global minimum. In addition, we fit the data using different initial conditions to confirm that the best fitted solution corresponds to the global minimum solution.
These algorithms can be used  to fit absolute astrometric data (e.g., planetary systems), relative astrometric data (e.g., binary systems), and combined (absolute plus relative) astrometric data (e.g., a planetary companion associated to a star in a binary system). To fit the astrometric data, we model the barycentric two-dimensional position of the source as a function of time ($\alpha(t)$, $\delta(t)$), accounting for the (secular) effects of proper motions ($\mu_\alpha$, $\mu_\delta$), accelerations terms ($a_\alpha$, $a_\delta$) due to a possible undetected companion with very long orbital period, the (periodic) effect of the parallax $\Pi$, and the (Keplerian) gravitational perturbation induced on the host star by one or more companions, such as low-mass stars, substellar companions, or planets (mutual interactions between companions are not taken into account). We searched for the best possible model (a.k.a, the closest fit) for a discrete set of observed data points ($\alpha(i)$, $\delta(i)$). The fitted function has several adjustable parameters, whose values are obtained by minimizing a $''$merit function$''$, which measures the agreement between the observed data and the model function. We minimize the $\chi^{2}$ function to obtain the maximum-likelihood estimate of the model parameters that are being fitted \citep[e.g.,][]{curiel19, curiel20}.

We also use the recursive least-squares periodogram method with a circular orbit (RLSCP) presented by \citet[][]{curiel19, curiel20} to search for astrometric signals that indicate the presence of possible companions. We start the search by comparing the least-squares fit of the basic model (proper motions and parallax only) and a one-companion model (proper motions, parallax, and Keplerian orbit of a single companion). If a signal is found, and it is confirmed by the two astrometric fitting models, we remove this signal by comparing the least-squares fits of a one- and a two-companion model (proper motions, parallax, and Keplerian orbits of one and two companions, respectively), and so on.
The RLSCP periodogram  follow a Fisher F$-$distribution with $k_{p} – k_{0}$ and $N_{obs} – k_{p}$ degrees of freedom, where $k_{p}$ and $k_{0}$ are the number of parameters that are fitted when the model includes a planetary companion and without a planetary companion, respectively. $N_{obs}$ is the number of observed epochs. Thus, an astrometric signal found in the periodogram indicates a high probability that a planetary companion is orbiting the star.

In addition, we also use the open-source package {\tt lmfit} \citep[][]{newville20}, which uses a non-linear least-squares minimization algorithm to search for the best fit of the observed data. This python package is based on the {\tt scipy.optimize} library \citep[][]{newville20}, and includes several classes of methods for curve fitting, including Levenberg-Marquardt minimization and  emcee  \citep[][]{foremanmackey13}. In addition, {\tt lmfit} includes methods to calculate confidence intervals for exploring minimization problems where the approximation of estimating parameter uncertainties from the covariance matrix is questionable.

\section{Results} \label{sec:results}

By combining our new data with the VLBA data from the archive \citep[][]{bower09,bower11}, we were able to search for sub-stellar companions associated to the main star GJ~896A in this binary system. 
These multi-epoch astrometric observations covered about 5317 days (16.56 yr), with an observational cadence that varies during the time observed. The observations were not spread regularly over the 16.56 years, the gaps between observations ranged from weekly to monthly to 7 years (see Sec.~\ref{sec:obs}).
There are a total of 16 epochs in the analysis presented here. The time span and cadence of the observations are more than adequate to fit the proper motions and the parallax of this binary system, and to search for possible companions with orbital periods between several days and a few years. 
In the following sections we present the resulting astrometric parameters for the fit with no companions (Sec.~\ref{sec:ssa}), a fit including a single (new planetary) companion (Sec.~\ref{sec:sca}), the fit of the relative astrometry of the binary (Sec.~\ref{sec:bsys}), a fit combining the absolute and relative astrometry (Sec.~\ref{sec:caf}), and finally a fit combining the absolute and relative astrometry plus a planetary companion (Sec.~\ref{sec:fcaf}).

\subsection{Single-source Astrometry } \label{sec:ssa}

\begin{figure*}
\centering
\plotone{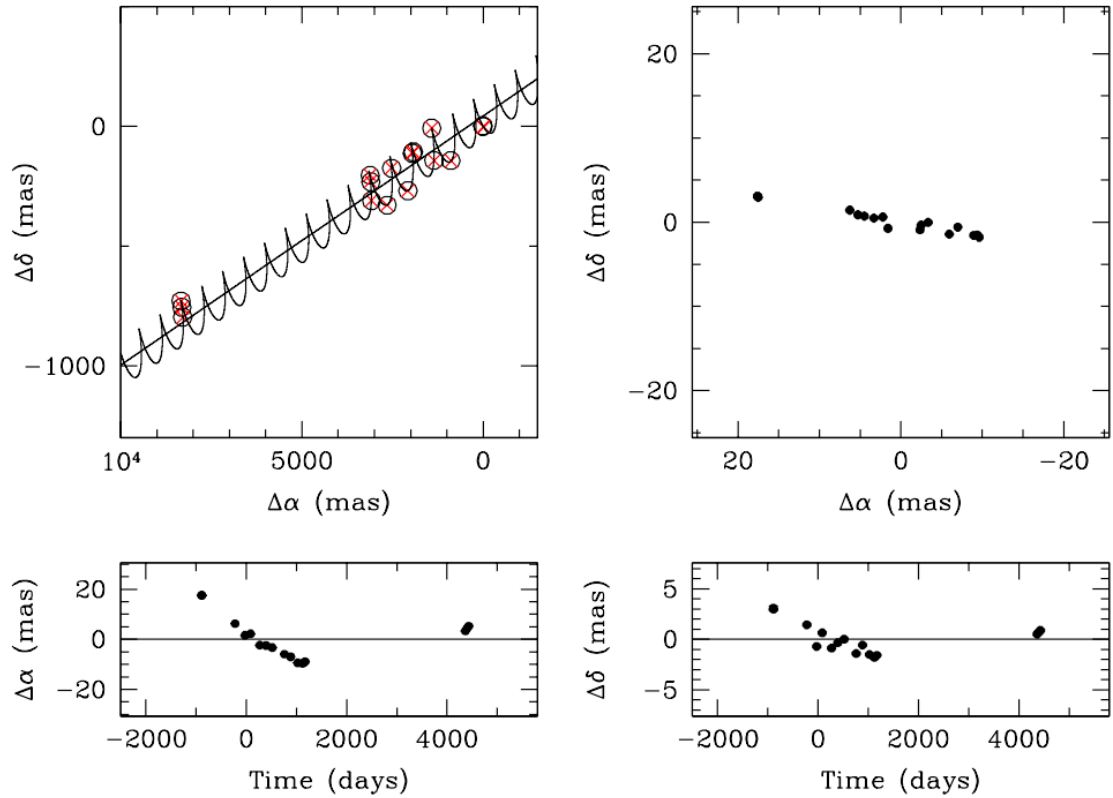}
    \caption{Absolute astrometry of the M Dwarf  GJ~896A using the VLBA observations and including only the parallax and proper motions in the fit. The upper left panel shows the observed data and the astrometric fit obtained when fitting only the proper motions and parallax of the star.  The upper right and lower panels show large residuals, having a well defined long term temporal trend.}   
    \label{fig_2}%
    \end{figure*}

First, both the least-squares and the AGA algorithms were used to fit the proper motions and the parallax of GJ~896A, without taking into account any possible companion. The results of the absolute astrometric fit are shown in column 1 of Table~\ref{tab_2} and Figure~\ref{fig_2}. We find that the residuals are quite large, and present an extended temporal trend that indicates the presence of a companion with an orbital period larger than the time span of the observations. 

The recursive least-squares periodogram with a circular orbit (RLSCP) of the astrometric data (see Figure~\ref{fig_3}, top panel)  shows a very strong signal that extent beyond the extend of the plot. 
A blind search for the orbital period of the signal indicates that the orbital period of this astrometric signal could not be constrained. 
This suggests that the signal may be due  to a companion with an orbital period much larger than the time span of the observations ($\gg$ 16 yr).  As we will see below (see Secs.~\ref{sec:caf} and \ref{sec:fcaf}), this long term signal is due to the gravitational interaction of GJ~896A with its very low-mass stellar companion GJ~896B. Since the orbital period of the binary system ($>$~200 yrs) is much larger than the time span of the VLBA observations ($\sim$16 yrs), the variation trend in the residuals can be taken into account by using accelerations terms in the astrometric fit.

The astrometric data was then fitted with the least-squares and AGA algorithms, including acceleration terms, which take into account the astrometric signature due to the low-mass stellar companion (single-source solution). The results of this single-source solution are shown in column 2 of  Table~\ref{tab_2} and Figures~\ref{fig_4}a. 
The astrometric solution shows that GJ~896A has proper motions of $\mu_\alpha$ = 574.171 $\pm$ 0.014 mas yr$^{-1}$ and $\mu_\delta$ = $-$60.333 $\pm$ 0.014 mas yr$^{-1}$. The fitted parallax is $\Pi$ = 160.027 $\pm$ 0.094 mas, which corresponds to a distance of 6.2489 $\pm$ 0.0037 pc. These values are similar to those obtained from the astrometric fit where no acceleration terms were taken into account.
The fitted acceleration terms are relatively large ($a_{\alpha}$ = 0.8893 $\pm$ 0.0034 mas yr$^{-2}$ and $a_{\delta}$ = 0.1402 $\pm$ 0.0035 mas yr$^{-2}$). We obtain now a better fit to the astrometric data, with smaller residuals (rms $\sim$ 0.24 mas in both R.A. and Dec.; see column 2 of Table~\ref{tab_2}). The total residuals (rms $\sim$ 0.34) are a factor of 25 smaller than those obtained with the astrometric fit without acceleration terms (rms $\sim$ 8.64). 
However, the residuals are large compared to the mean noise present in the data (rms $\sim$0.11 and 0.10 mas for R.A. and Dec., respectively) and the astrometric precision expected with the VLBA ($<$70 $\mu$as). Below we investigate the possibility that these residuals are due to a planetary companion orbiting around GJ~896A.

\subsection{Single-companion Astrometry } \label{sec:sca}

\begin{figure*}
   \centering
   \plotone{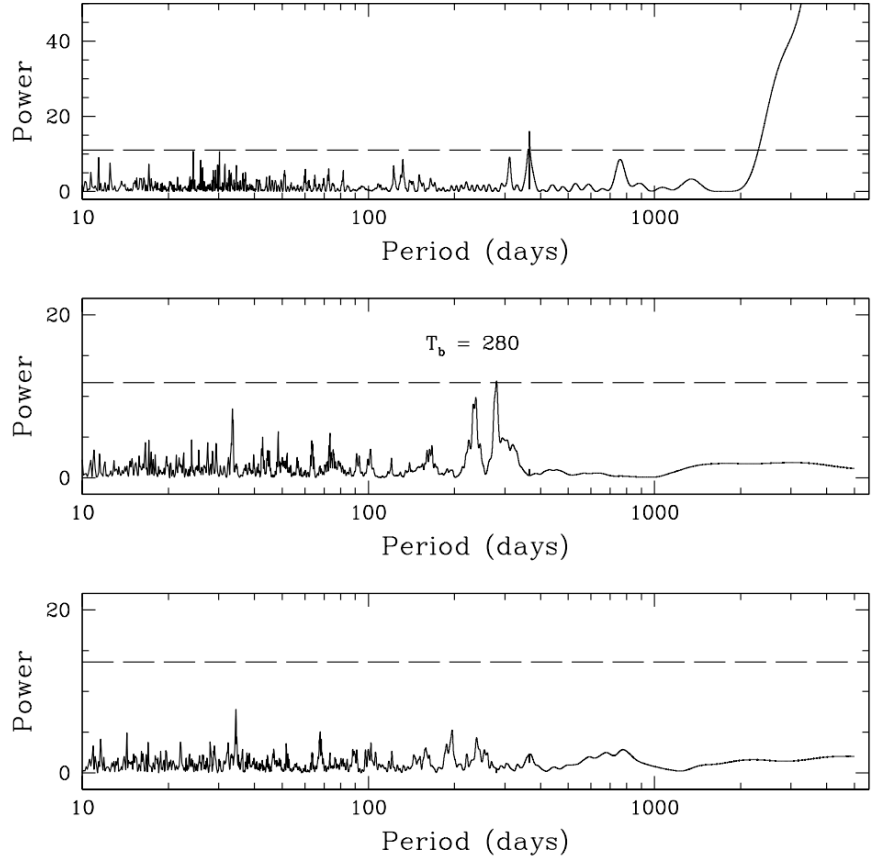}
    \caption{RLSCP periodograms obtained by fixing the eccentricity $e = 0$. The upper panel shows the periodogram obtained with the astrometric VLBA data of the star GJ~896A, including only the parallax and proper motions in the fit. This periodogram clearly shows a prominent signal that goes beyond the time interval of the plot. This signal cannot be restricted with the present absolute astrometric data (see Secs.~\ref{sec:ssa} and \ref{sec:sca}). The middle panel shows  the periodogram when adding acceleration terms to the absolute astrometric fit, that take into account the very long term signal indicated in the upper periodogram. One strong signal appears in this periodogram with an orbital period of about 280 days and with a false alarm probability of FAP = 0.9\%. The lower panel shows  the periodogram  obtained by including acceleration terms and removing a signal of 280 days. No clear signals are observed in this periodogram.  The horizontal line indicates the limit of the false alarm probabilities FAP = 1\%.}
    \label{fig_3}%
    \end{figure*}

\begin{figure*}
\centering
\gridline{
              \fig{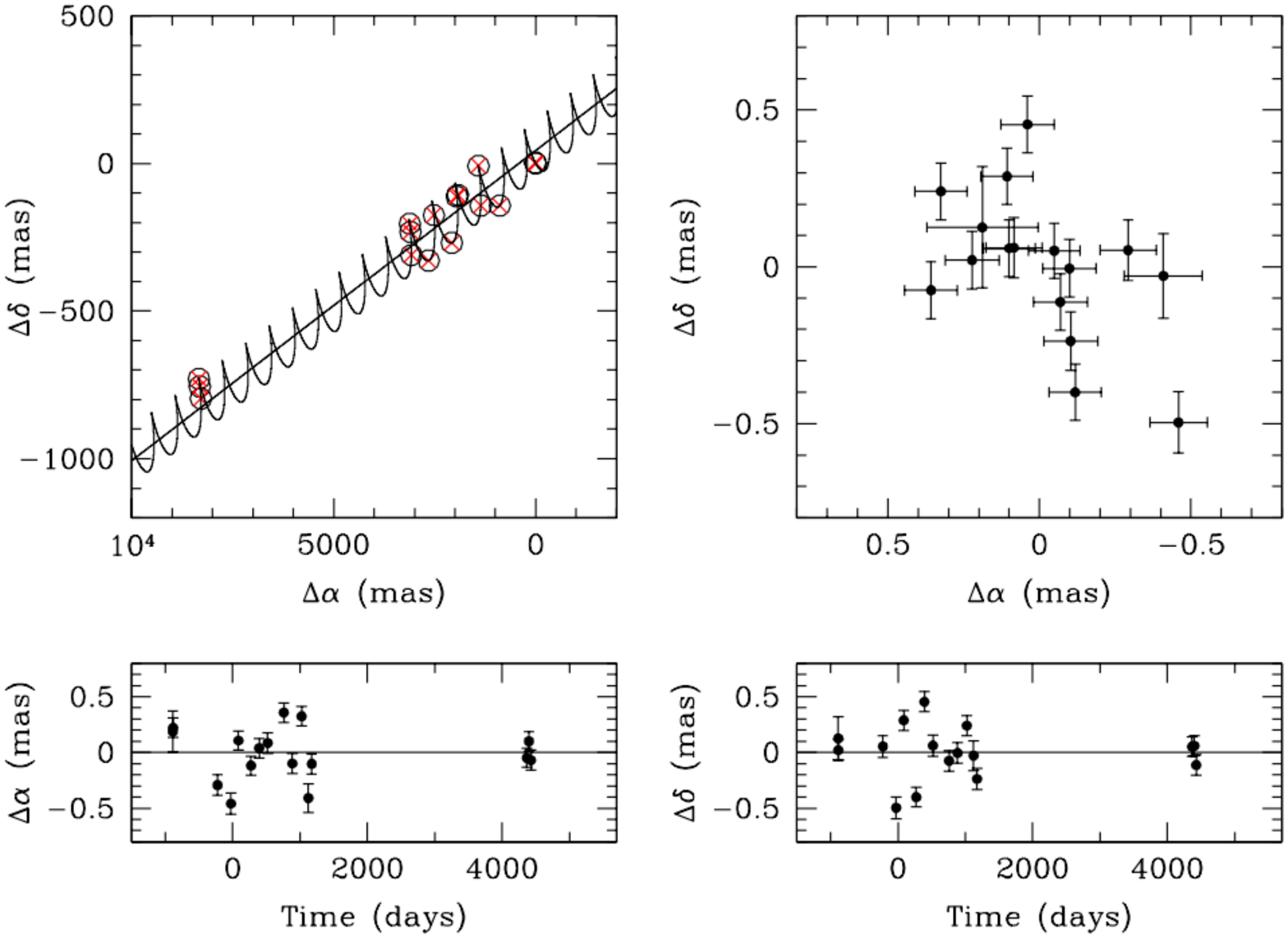}{0.48\textwidth}{a}
              \fig{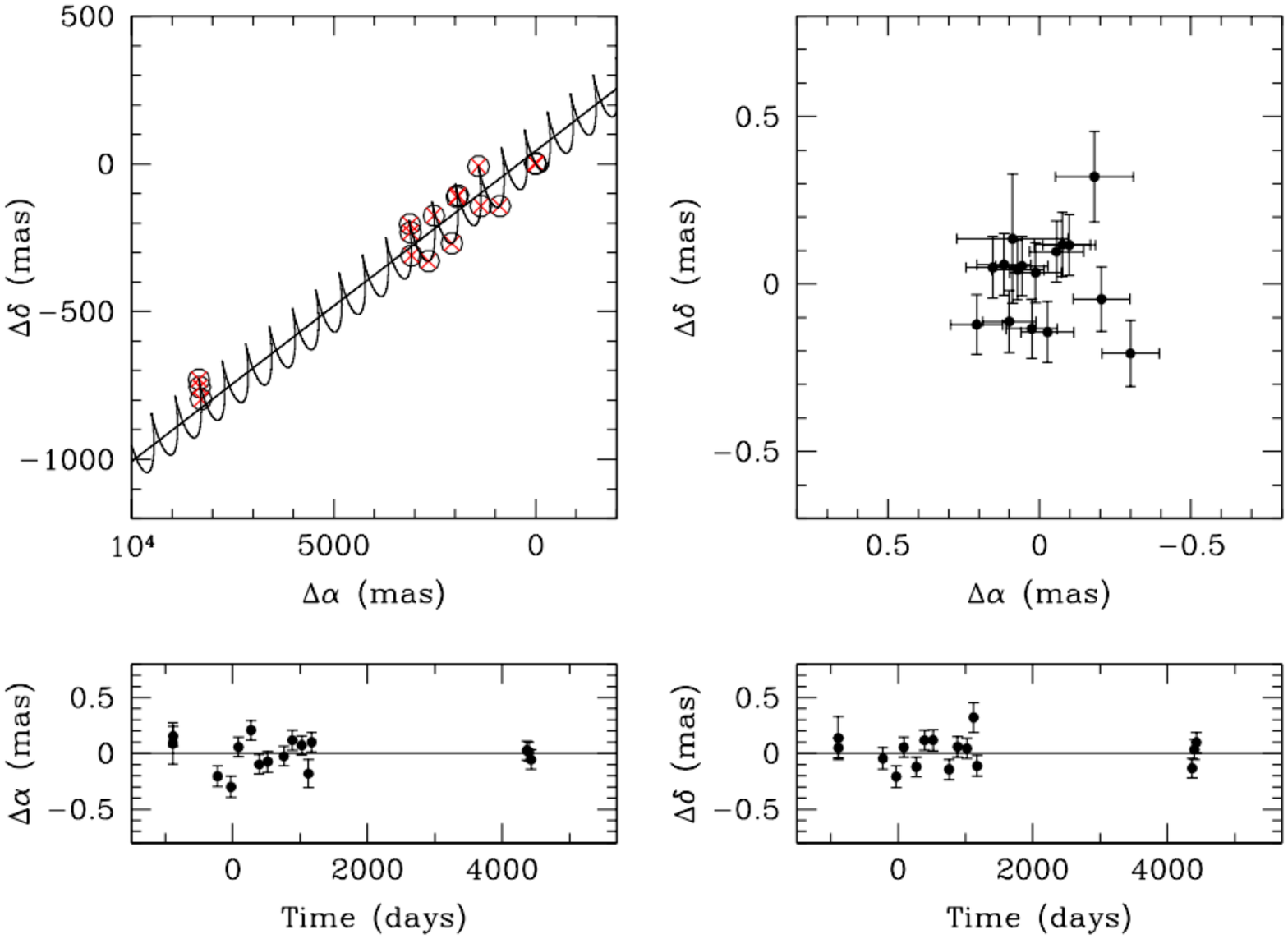}{0.48\textwidth}{b}
             }
    \caption{{\bf (a)} Absolute astrometry of the M Dwarf  GJ~896A, including only acceleration terms in the fit. The residuals are smaller than those obtained from the single-source solution, where accelerations terms are not included. The residuals, shown in upper-right and lower panels, are still larger than the astrometric precision of the observations. {\bf (b)} Similar to the left figure, but in this case the panels show the absolute astrometry of the M Dwarf  GJ~896A, including acceleration terms and the astrometric signal of the star due to a planetary companion. The residuals are smaller than those shown in {\bf (a)}, but they are still larger than the astrometric precision of the observations.}   
    \label{fig_4}%
    \end{figure*}

\begin{figure*}
\centering
\gridline{
              \fig{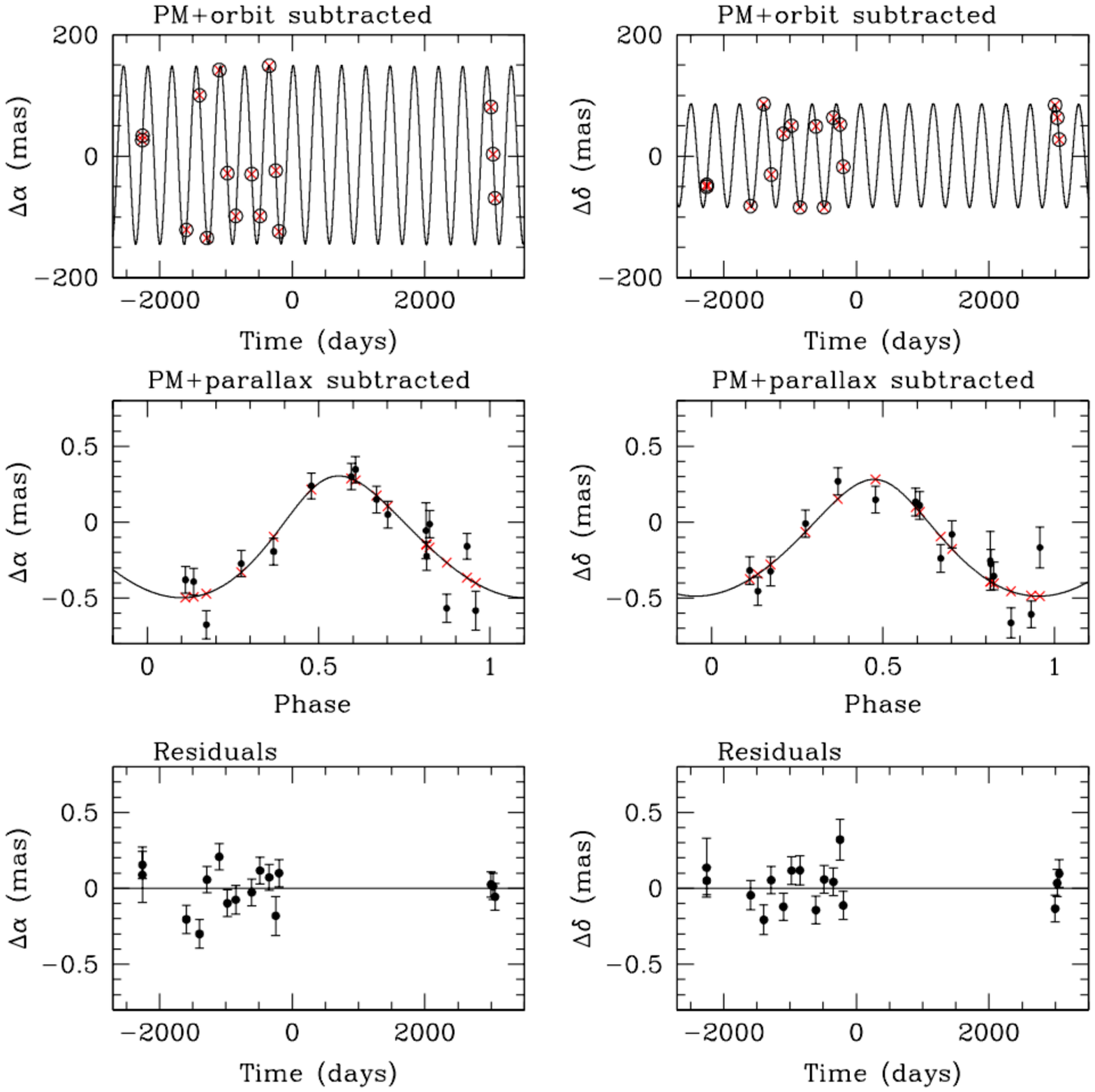}{0.48\textwidth}{a}
              \fig{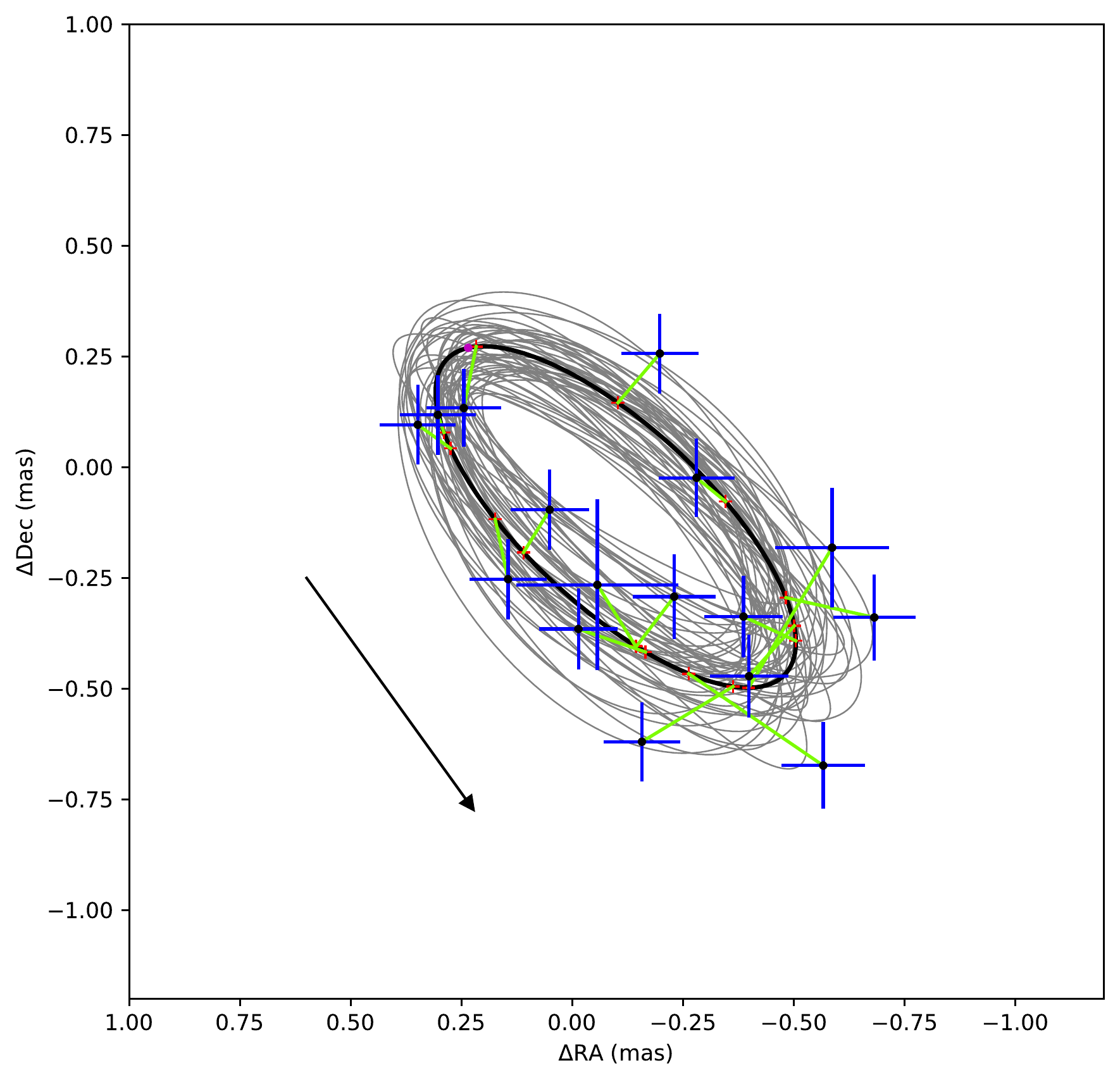}{0.48\textwidth}{b}
             }
    \caption{{\bf (a)} Single-companion astrometric fit of GJ~896A using only the VLBA data. The upper panels show the parallax fit of the source after subtracting proper motions and the astrometric signal of the star due to the planetary companion. The middle panels show the astrometric fit of the main star after removing parallax and proper motions. The lower panels show the residuals of the astrometric fit, which are the same shown in Fig.~\ref{fig_5}b.  {\bf (b)} Orbital motion of the main source GJ~896A due to gravitational pull of the planetary companion. The crosses indicate the observational data, and the green arrows connect the observed position with the expected location on the orbit at each observed epoch, which is indicated with a red cross. The magenta dot indicates the position of the periastron. 
Shown in light gray are fifty orbits generated randomly from the planet's orbital parameters (reported in Table \ref{tab_2}) within one sigma.
The long arrow indicates the direction of the orbital motion.}   
    \label{fig_5}%
    \end{figure*}

We carried out a RLSCP of the astrometric data, including accelerations terms. The RLSCP shows now at least two significant peaks (see Figure~\ref{fig_3}, middle panel), the most prominent one with a period of about 280 days, and the weaker signal with a period of about 237 days. The stronger signal appears to be well constrained, with a false alarm probability (FAP) of about 0.90\%. This  low FAP suggests that the main signal in the periodogram is real and due to the presence of a companion in a compact orbit. 
We then used both the  least-squares and AGA algorithms to fit the astrometric observations of GJ~896A, now including acceleration terms and a possible single companion in a compact orbit (single-companion solution). The solution of the fit is shown in column 3 of Table~\ref{tab_2} and Figures~\ref{fig_4}b and \ref{fig_5}. The fit of the astrometric data clearly improves when including a companion, as seen by the $\chi^{2}_{red}$. The $\chi^{2}_{red}$ is now a factor 2.4 smaller than that obtained with the single-source solution. Table~\ref{tab_2} and Figure~\ref{fig_4} show that the residuals of the single-companion solution (rms $\sim$0.14) are a factor of 1.7 smaller than in the case of the single-source solution (rms $\sim$0.24). 

Column 3 in Table~\ref{tab_2} summarizes the best fit of the VLBA astrometric data, including this new companion in a compact orbit. 
The single-companion solution shows that GJ~896A has proper motions of $\mu_\alpha$ = 574.142 $\pm$ 0.019 mas yr$^{-1}$ and $\mu_\delta$ = $-$60.354 $\pm$ 0.020 mas yr$^{-1}$. The fitted parallax is $\Pi$ = 159.83 $\pm$ 0.13 mas, which corresponds to a distance of 6.2565 $\pm$ 0.0051 pc.
The fitted acceleration terms are $a_{\alpha}$ = 0.8871 $\pm$ 0.0047 mas yr$^{-2}$ and $a_{\delta}$ = 0.1400 $\pm$ 0.0049 mas yr$^{-2}$.
These values are similar to those obtained from the single-source solution plus acceleration.

 Figure~\ref{fig_5} shows the orbital motion of the star GJ~896A due to the gravitational pull of the companion.
The orbit of the main star around the barycenter has an orbital period $P$ = 281.56 $\pm$ 1.67 days, an eccentricity $e = 0.30 \pm 0.11$, a longitude of the periastron $\omega = 344.0^\circ \pm 13.1^\circ$, a position angle of the line of nodes $\Omega = 47.7^\circ \pm 12.1^\circ$, a semi-major axis $a_{A} = 0.52 \pm 0.11$ mas, and an inclination angle $i = 66.0^\circ \pm 15.0^\circ$, which indicates that the orbit is prograde ($i < 90^\circ$). The orbit of the star around the barycenter due to the companion is well constrained, which is consistent with the narrow signal observed in the periodogram (see Figure~\ref{fig_3}, middle panel). With this astrometric fit alone we can not estimate the mass of the companion. However, we used the estimated mass of the planetary system GJ~896A that we obtain below, using the full combined astrometric fit of this binary system (see Sec.~\ref{sec:fcaf}). Column 3 in Table~\ref{tab_2} summarizes the parameters of the new companion, here after GJ~896A$b$. We find that its mass is 0.00223 $\pm$ 0.00047 M$_\odot$, which is consistent with a planetary companion with a mass of 2.35 $\pm$ 0.49 M$_{J}$. The orbit of the planetary companion around the barycenter  has a semi-major axis $a_{b}$ = 0.6352 $\pm$ 0.0018 au (or 101.53 $\pm$ 0.42 mas).

There is an ambiguity with the position angle of the line of nodes between $\Omega$ and  $\Omega + 180^{\circ}$. This ambiguity can be solved by radial velocity (RV) observations. However, since this planetary companion has not been detected with RV observations, we will leave the fitted $\Omega$  angle in Table~\ref{tab_2}. We further discuss this ambiguity below (see section~\ref{sec:mutinc}).

The orbit of the planetary companion is well constrained. However, the rms of the residuals ($\sim$0.14 mas) are still large compared to the mean noise present in the data and the astrometric precision expected with the VLBA. 
To investigate the possibility of a second possible planetary companion, we obtained a new RLSCP, including acceleration terms and a signal with a period of 281.6 days. The new RLSCP does not show a significant signal (see Figures~\ref{fig_3}, bottom panel). 
We also carried out a blind search for a second planetary companion using the least-squares and AGA algorithms. We did not find a possible candidate. We will need further observations to investigate whether or not this source has more planetary companions. 

\subsection{Binary-system Astrometry: Relative fit} \label{sec:bsys}

GJ~896AB is a visual binary that contains two M dwarf stars that has been observed in the optical and near infrared for more than 80 yr. 
The Washington Double Star \citep[WDS,][]{mason01} Catalog provides separation and position angle measurements of this binary system spanning these decades and go back as far as 1941. Since the stellar companion  GJ~896B was detected in two of our recent VLBA observations of this binary system, we include the separation  and the position angle of these two epochs in the relative astrometric fit. We performed an  astrometric fit to the relative astrometric measurements with both, the least-squares and the AGA algorithms (binary-system solution). Since the WDS catalog does not provide estimated error bars for most of the observed epochs, we applied a null weight to the astrometric fit, which translate into a uniform weight for all the observed epochs. 

\begin{figure*}
   \centering
    \plotone{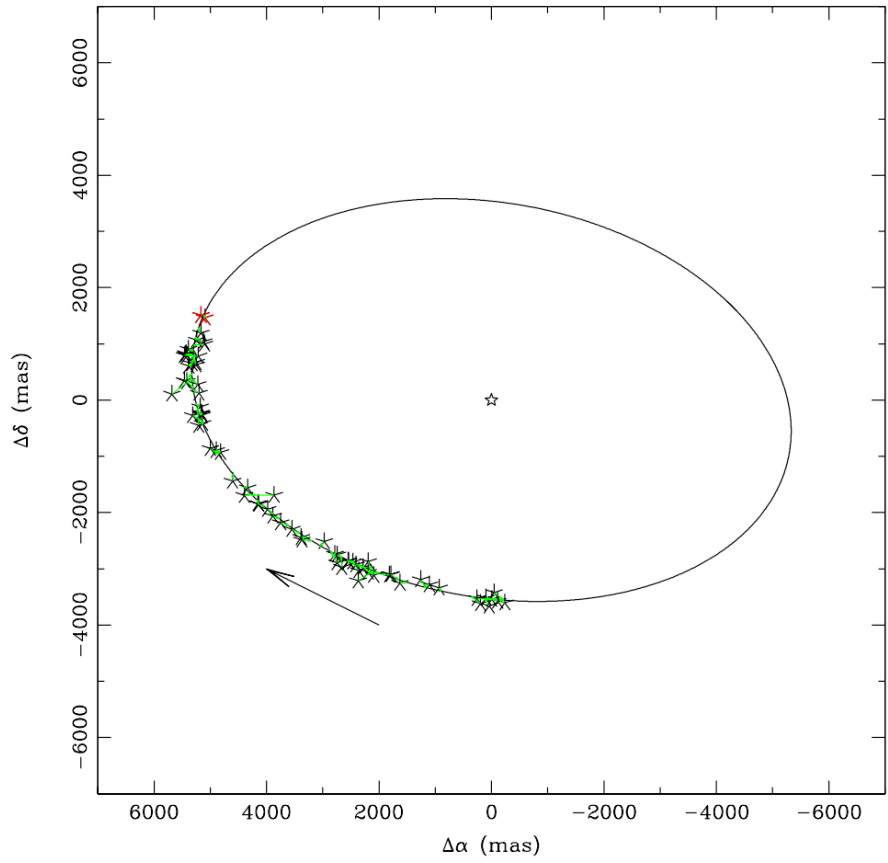}
    \caption{This plot shows the orbital motion of the very low mass star GJ~896B around the main star GJ~896A. The stars correspond to the relative astrometry measurements of the binary system GJ~896AB using optical and infrared observations (black stars), as well as the radio detection of both stars (red stars). The green arrows connect the observed data with the location on the orbit of the fitted position at each observed epoch. The star at the center indicates the position of the main source GJ~896A. The long arrow indicates the direction of the orbital motion.}   
    \label{fig_6}%
    \end{figure*}

The results of the binary-system solution are presented in column 1 of Table~\ref{tab_3} and Figure~\ref{fig_6}. We find that the relative astrometric fit of this binary system is well constrained when assuming a circular orbit. When the eccentricity of the binary system is taken into account, the solution does not converge to a stable solution. The orbit of the low-mass stellar companion GJ~896B around the main star  GJ~896A has an orbital period $P_{AB}$ = 96606.13 $\pm$ 2.24 days (264.5 yr), a position angle of the line of nodes $\Omega_{AB} = 79.7083^\circ \pm 0.0054^\circ$, a semi-major axis $a_{AB} = 5378.79 \pm 0.51$ mas, and an inclination angle $i_{AB} = 130.6592^\circ \pm 0.0087^\circ$, which indicates that the orbit of the stellar companion is retrograde ($i > 90^\circ$). Using the distance to the source that we obtain from the full combined astrometric fit (see Sec.~\ref{sec:fcaf}), we obtain that the mass and the semi-major axis of this binary system are $m_{AB}$ = 0.544304 M$_\odot$ and a$_{AB}$ = 33.64265 au, respectively.

As it was mentioned before, there is an ambiguity with the position angle of the line of nodes between $\Omega$ and  $\Omega + 180^{\circ}$. In this case, the observed RV of the two M dwarfs can be used to find the correct angle.
From the GAIA catalog \citep[][]{lindegren18}, the observed mean radial velocity of GJ~896A is $-0.02 \pm 0.31$ km s$^{-1}$. The GAIA catalog does not contain the radial velocity of GJ~896B. However, recent RV observations of this source indicate that its RV is about 3.34 km s$^{-1}$ \citep[][]{morin08}. This means that GJ~896B is red shifted with respect to GJ~896A. Thus, the position angle of the line of nodes that is consistent with these RVs is $\Omega_{AB} = 259.7083^\circ \pm 0.0054^\circ$. 
This correction to the position angle of the line of nodes is applied in all the astrometric fits presented in Table~\ref{tab_3}.

\subsection{Binary-system Astrometry: Combined Astrometric fit} \label{sec:caf}

The optical/infrared relative astrometry of the binary system GJ~896AB  \citep[][]{mason01}, obtained in a time interval of about 80 yrs, was also combined with the absolute radio astrometry to simultaneously fit: the orbital motion of the two stars in GJ~896AB; the orbital motion of the planetary companion around GJ~896A (see Sec.~\ref{sec:fcaf}); and the parallax and proper motion of the entire system. However, since most of the relative astrometric observations do not have estimated errors, for the relative astrometry part we carried out a uniform weighted fit (without errors) of the observed data (73 epochs, including the two relative positions of GJ~896B obtained with the VLBA), while for the absolute astrometry of the VLBA observations of the main star GJ~896A (16 epochs) and the secondary star GJ~896B (2 epochs), we applied a weighted fit using the estimated error of the observed position at each epoch. 
The results presented here and in Sec.~\ref{sec:fcaf} were obtained by fitting simultaneously the absolute astrometric observations of both stars, and the relative astrometry of this binary system.
The results are limited by: a) the lack of error bars of the relative astrometric data, b) the relative astrometric data covers about 35$\%$ of the orbital period of the binary system ($\sim$229.06 yrs; see Sec.~\ref{sec:fcaf}), and c) the absolute astrometry covers only a time span of about 16.56 yrs. Thus, all errors reported below are statistical only.

Fitting the proper motions and parallax of a binary system is complex due to the orbital motions of each component around the center of mass, especially since both stars have  different masses ($m_{B}/m_{A} <$ 1). The proper motion and parallax of a binary system can be obtained with high precision when the orbital period of the system is small compared with the time span of the astrometric observations. In contrast, when the time span of the astrometric observations is small compared with the orbital period of the binary system, it is difficult to separate the orbital motion of each component and the proper motion of the system since both spatial movements are blended. The best way to separate both movements is to simultaneously fit the proper motions, the parallax, and the orbital motion of the binary system. Our combined astrometric fit includes all these components, and thus gives an accurate estimate of the proper motions and the parallax of this binary system.

We carried out a combined fit of the relative and absolute astrometric data and found that the astrometric fit improves substantially. The solution of the combined astrometric fit is similar to that obtained from the relative astrometric fit. However, using the combined astrometric fit, we are able to obtain the parallax and the proper motions of the binary system, as well as improved orbits of the two stars around the center of mass and the masses of the system and the individual stars.
The results of this combined fit are shown in Figures~\ref{fig_7}a and \ref{fig_8}a, and summarized in column 2 of Table~\ref{tab_3}. 
Although, Figure~\ref{fig_8}a shows the full combined astrometric solution (including a planetary companion; see Sec~\ref{sec:fcaf}), the solution for the relative astrometric part of the binary system is basically the same as that obtained from the combined astrometric solution (see columns 2 and 3 of Table~\ref{tab_3}). Thus, we present only one figure showing both fits.
In contrast to the case of the relative astrometric fit (see~Sec.~\ref{sec:bsys}), we find here that the combined astrometric fit is well constrained when considering that the orbit of the binary system could be eccentric (e $\ge$ 0). The proper motions of the binary system are $\mu_\alpha$ = 571.515 $\pm$ 0.019 mas yr$^{-1}$ and $\mu_\delta$ = $-$37.750 $\pm$ 0.019 mas yr$^{-1}$. The parallax of the binary system is $\Pi$ = 159.98 $\pm$ 0.14 mas, which corresponds to a distance of 6.2506 $\pm$ 0.0055 pc. The orbit of the binary system has an orbital period $P_{AB}$ = 83665.80 $\pm$ 1.64 days (229.06 yr), a position angle of the line of nodes $\Omega_{AB} = 255.0891^\circ \pm 0.0028^\circ$, a semi-major axis of the binary system $a_{AB} = 5057.96 \pm 0.36$ mas,  a semi-major axis of the main source $a_{AB}(A) = 1381.015 \pm 0.069$ mas, and an inclination angle $i_{AB} = 130.0664^\circ \pm 0.0085^\circ$. 
The estimated mases of the binary system and the two components are $m_{AB}$ = 0.602253 $\pm$ 0.000020 M$_\odot$, $m_{AB}(A)$ = 0.43782 $\pm$ 0.00054 M$_\odot$, and $m_{AB}(B)$ = 0.16444 $\pm$ 0.00020 M$_\odot$, respectively. The semi-major axis of the binary system and the two stellar components are $a_{AB}$ = 31.615 $\pm$ 0.028, $a_{AB}(A)$ = 8.6322 $\pm$ 0.0075, and $a_{AB}(B)$ = 22.983 $\pm$ 0.020 au, respectively.

Column 2 of Table~\ref{tab_3}  indicates that there is a relatively small improvement of about 17\% in the $\chi^{2}_{red}$. This is because the residuals of the relative data of the binary system are much larger that those of the absolute astrometry of the main star, and this dominates the residuals.
However, Table~\ref{tab_3} and Figure~\ref{fig_7} show that the residuals of the relative astrometric data of the binary system from the combined astrometric solution (rms $\sim$116.38 mas) are a factor of 1.3 smaller than in the case of the relative astrometry  solution (rms $\sim$152.0 mas).
Thus, the combined astrometric solution is an improvement to the solution obtained from the pure relative astrometric fit.
In addition, the residuals of the absolute astrometric data of the main source GJ~896A (rms $\sim$ 0.27 and 0.47 mas for R.A. and Dec.) are large compared to the mean noise present in the data. 
 
Comparing the residuals that were obtained from the single-source fit (see~Sec.~\ref{sec:ssa}) of the astrometric observations of the main star GJ~896A, including acceleration terms, and the residuals of the combined astrometric fit (see column 2 of Table~\ref{tab_2} and column 2 of Table~\ref{tab_3}, and Figures~\ref{fig_4} and \ref{fig_7}), we find that the rms of the residuals of the absolute astrometry are similar. In addition, we find that the temporal distribution of the residuals are also very similar. However, the residuals of the last three epochs from the  combined astrometric fit are considerable larger that those obtained from the single-source plus acceleration fit. 
We do not find an explanation for this discrepancy.
But, we notice that two out of these three epochs are the ones where the secondary star was detected, and the coordinates of this star were used in the fit of the relative astrometry of the binary system and in the absolute astrometry fit of the secondary star GJ~896B. Since the secondary star seems to be a close binary (see Secs.~\ref{sec:intro} and ~\ref{sec:pmot}) and it was only detected in two epochs, this probably affects the astrometric fit (see Figure~\ref{fig_7}). Further observations will be needed to improve the astrometric fit of this binary system. 

As an independent check of the derived parameters, we performed a combined fit of the astrometric data with the {\tt lmfit} package \citep[][]{newville20}, which uses a non-linear least-squares minimization algorithm to search for the best fit of the observed data  (see~Appendix~\ref{sec:mcmc} for further details). We find that  the astrometric fit is well constrained, and that the solution and the residuals of the fit are in good agreement with those derived with the combined astrometric fit (see~Sec.~\ref{sec:caf}).

These results suggest that the main star may have at least one companion. Below we investigate further the possibility that these residuals are due to a planetary companion orbiting around the main star GJ~896A (as in the case of the single-companion fit, see Sec.~\ref{sec:sca}).

\subsection{Binary-system plus planet Astrometry: Full Combined Astrometric fit} \label{sec:fcaf}

We carried out a combined astrometric fit of the relative orbit of the binary and the absolute astrometric data of the main and the secondary stars, now including the orbital motion of a possible companion in a compact orbit (full combined astrometric fit), and found that the astrometric fit improves substantially. The results of this full combined astrometric fit are presented in Figures~\ref{fig_7}b, \ref{fig_8} and \ref{fig_9}. We find that the full combined astrometric fit is well constrained. Column 3 of Table~\ref{tab_3} summarizes the best fit parameters of the full combined astrometric fit.

By combining relative and absolute astrometric data of the binary system, we are able to obtain the full dynamical motion of the system, including the close companion.
From the best fit of the full combined astrometric data, we obtained proper motions $\mu_\alpha$ = 571.467$\pm$0.023 mas yr$^{-1}$ and $\mu_\delta$ = $-$37.715$\pm$0.023 mas yr$^{-1}$, and a parallax $\Pi$ = 159.88$\pm$0.17 mas for the binary system (see column 3 of Table~\ref{tab_3}). The estimated proper motions and parallax are very similar to those obtaned from the Combined astrometric fit. However,  as expected, the proper motions differ from those obtained by GAIA for both stars GJ~896A and GJ~896B, which where obtained from independent linear fits of both stars \citep[][]{lindegren18}. The parallax, which corresponds to a distance of $d$ = 6.2545$\pm$0.0066 pc, is an improvement to that obtained by GAIA for both stars ($\Pi_{A}$ = 159.663$\pm$0.034 mas, and $\Pi_{B}$ = 159.908$\pm$0.051 mas). The parallax estimated by us is for the binary system, while GAIA's parallaxes are for the individual stars.

The part of the solution that corresponds to the astrometric fit of the binary system is very similar to that obtained from the combined astrometric fit (see columns 2 and 3 of Table~\ref{tab_3}). The fit of the binary system does not improve in this case.
However, the best full combined fit of the astrometric data indicates that the M3.5 dwarf  GJ~896A has at least one planetary companion, GJ~896A$b$. 
The orbit of the star around the barycenter, due to this planetary companion, has an  orbital period $P_{Ab}$ = 284.39 $\pm$ 1.47 days, an eccentricity $e_{Ab} = 0.35 \pm 0.19$, a longitude of the periastron $\omega_{A} = 353.11^\circ \pm 11.81^\circ$, a position angle of the line of nodes $\Omega_{Ab} = 45.62^\circ \pm 9.60^\circ$, a semi-major axis $a_{A} = 0.51 \pm 0.15$ mas, and an inclination angle $i_{Ab} = 69.20^\circ \pm 25.61^\circ$, which indicates that the orbit is well constrained and prograde ($i_{Ab} < 90^\circ$).
However, the rms of the residuals ($\sim$0.21 and $\sim$0.45  mas in R.A and Dec., respectively) are still large compared to the mean noise present in the data ($\sim$0.15 mas) and the astrometric precision expected with the VLBA ($<$70 $\mu$as). This may explain the large errors of the orbital parameters. The large residuals may also indicate the presence of a second planetary companion. We carried out a blind search for a possible astrometric signal due a second planetary companion using the least-squares and the AGA algorithms, however, we did not find a candidate.

Using the fitted solution, we obtain the total mass of the system ($m_{AB}$ = 0.603410$\pm$0.000025 $M_{\odot}$), the masses of the two stars ($m_{AB}(A)$ = 0.43814$\pm$0.00065 $M_{\odot}$ and $m_{AB}(B)$ = 0.16527$\pm$0.00025 $M_{\odot}$). In addition, we find that the mass of the star GJ~896A is $m_{A}$ = 0.43599$\pm$0.00092 $M_{\odot}$, and that the mass of the planetary companion is $m_{b}$ = 0.00216$\pm$0.00064 $M_{\odot}$, which is consistent with being planetary in origin with an estimated mass of 2.26$\pm$0.57 $M_{J}$.
The semi-major axis of the orbit of the binary system is $a_{AB}$ = 31.635 $\pm$ 0.033 au. The semi-major axis of the two stars around the center of mass are $a_{AB}(A)$ = 8.6646 $\pm$ 0.0091 au and $a_{AB}(B)$ = 22.971 $\pm$ 0.024 au, and the semi-major axis of the orbit of the planetary companion is $a_{b}$ = 0.63965 $\pm$ 0.00067 au (or 102.27 $\pm$ 0.15 mas).

This is the first time that a planetary companion of one of the stars in a binary system has been found using the astrometry technique. The solution of the full combined astrometric fit of the planetary companion is similar to that obtained from the single-companion solution (see~Sec.~\ref{sec:sca} and Tables~\ref{tab_2} and \ref{tab_3}). 
The fact that the astrometric signal appears in the periodogram of the absolute astrometric observations of GJ~896A, and in both, the single-companion astrometric fit and the full combined astrometric fit obtained, in both cases with two different algorithms (least-squares and AGA), supports the detection of an astrometric signal due to a companion. Furthermore, the similar astrometric fit solution obtained from the single-companion astrometric fit and the full combined astrometric fit further supports the planetary origin of the astrometric signal.

\section{Discussion}  \label{sec:discusion}

\subsection{Proper Motions and Orbital Acceleration} \label{sec:pmot}

As mentioned before, the estimation of the proper motions of a binary system is complex due to the orbital motions of each component around the center of mass, specially when both stars have  different masses ($m_{B}/m_{A} <$ 1) and the time span of the observations cover only a fraction of the orbital period of the binary system. In such a case, it is difficult to separate the orbital motion of each component and the proper motion of the system, both movements are blended. The best way to separate both movements is to simultaneously fit the proper motions, the parallax, and the orbital motion of the binary system. The full combined astrometric fit that we carried out (see Sec.~\ref{sec:fcaf}) includes all these components, and thus gives good estimates of the proper motion and the orbital motion of the binary system (see Table~\ref{tab_3}). 

The full combined astrometric solution (see column 3 of Table~\ref{tab_3}) shows that the orbital motions of the two stars around the center of mass and of the planetary companion GJ~896A$b$ around the main star GJ~896A are well constrained. In addition, we found a similar solution for the orbital motion of the planetary companion using the full combined astrometric fit (see Sec.~\ref{sec:fcaf} and column 3 of Table~\ref{tab_3}) and the single-companion astrometric fit (see Sec.~\ref{sec:sca} and column 3 of Table~\ref{tab_2}). In the former case, the astrometric fit includes intrinsically the orbital motion of  GJ~896A around the center of mass of the binary system, while in the latter case, the included acceleration terms take into account this orbital motion. 

Tables~\ref{tab_2} and \ref{tab_3} show that the proper motions of the system obtained with the single-source,  single-companion and the full combined astrometric fits differ significantly. Here we discuss the origin of this difference.
The single-source astrometric solution gives $\mu_\alpha$ = 576.707 mas yr$^{-1}$ and $\mu_\delta$ = $-$59.973 mas yr$^{-1}$ (see Sec.~\ref{sec:ssa} and column 1 of Table~\ref{tab_2}), where we include only the proper motions and the parallax to the absolute astrometric fit of the main star GJ~896A. Including acceleration terms (single-source plus acceleration astrometric solution), to take into account the orbital motion of the binary system, the estimated proper motions of the system are $\mu_\alpha$ = 574.171 mas yr$^{-1}$ and $\mu_\delta$ = $-$60.333 mas yr$^{-1}$ (see Sec.~\ref{sec:sca} and column 2 of Table~\ref{tab_2}). 
On the other hand, from the full combined astrometric solution, the proper motions of the system are $\mu_\alpha$ = 571.467 mas yr$^{-1}$ and $\mu_\delta$ = $-$37.715 mas yr$^{-1}$ (see Sec.~\ref{sec:fcaf}  and column 3 of Table~\ref{tab_3}).
The absolute astrometric solutions presented in Table~\ref{tab_2} show that by including the acceleration terms in the fit, we improve the estimated proper motions of the system, and that the inclusion of an astrometric signal due to a companion in the fit does not affect the estimation of the proper motions and the estimated acceleration terms.
Comparing the single-source and the full combined astrometric solutions, we find a significative difference in the estimated proper motions, particularly in the north-south direction ($\Delta\mu_\alpha$ = 5.24 mas yr$^{-1}$ and $\Delta\mu_\delta$ = $-$22.26 mas yr$^{-1}$).
This is consistent with the fact that the VLBA astrometric observations of GJ~896A were obtained when the source was crossing the ascending node of the binary orbit and mainly in the north-south direction (see discussion below, and Figures~\ref{fig_8} and \ref{fig_9}).
These results indicate that the orbital motion of GJ~896A around the center of mass of the system blends with the proper motions of the binary system, and in second order with the acceleration terms included in the absolute astrometric fit. Thus, we conclude that the best estimates of the proper motions of this binary system are those obtained with the full combined astrometric fit (see Table~\ref{tab_3}).

We can also obtain an estimation of the velocity on the plane of the sky of the center of mass $\mu(CM)$ of this binary system, as follows:

\begin{equation}
\mu_{\alpha}(A) =  \mu_{\alpha}(CM) + v_{\alpha}(A), \\
\end{equation}
\begin{equation}
\mu_{\delta}(A) =  \mu_{\delta}(CM) + v_{\delta}(A),
\end{equation}
\noindent
and
\begin{equation}
v_{\alpha}(A) =  (2 \pi a_{A}/P_{AB}) \times sin(\theta_{rot}) \times cos(\phi),
\end{equation}
\begin{equation}
v_{\delta}(A) =  (2 \pi a_{A}/P_{AB}) \times cos(\theta_{rot}) \times cos(\phi),
\end{equation}

\noindent
where $\mu_{\alpha}(A)$ and  $\mu_{\delta}(A)$ are the observed proper motion of the main source GJ~896A (single-source solution; see column 1 of Table~\ref{tab_2}), $\mu_{\alpha}(CM)$ and $\mu_{\delta}(CM)$  are the proper motions of the center of mass of the binary system,  $v_{\alpha}(A)$ and  $v_{\delta}(A)$ are  the projected rotational velocity of GJ~896A on its orbital motion around the center of mass, $a_{A}$ is semi-major axis of the orbital motion of GJ~896A around the center of mass, P$_{AB}$ is the orbital period of the binary system, $\theta_{rot}$ $\approx$  $\Omega_{AB} - 90^\circ$ is the position angle of the projected rotation velocity (tangential velocity) of GJ~896A around the center of mass of the system (see Figures~\ref{fig_8}b and \ref{fig_9}), $\phi = 180^{\circ} - i_{AB}$ is the complementary angle of the orbital plane of the binary system, and $ i_{AB}$ is the fitted inclination angle of the orbital plane of the binary system (see column 3 of Table~\ref{tab_3}). Using the fitted parameters presented in column 1 of Table~\ref{tab_2} and column 3 of  Table~\ref{tab_3}, $\mu_{\alpha}(A)$ = 576.707 mas yr$^{-1}$ and  $\mu_{\delta}(A)$ = $-$59.973 mas yr$^{-1}$ ($\mu_{A}$ = 579.8170 mas yr$^{-1}$ with PA = 95.94$^\circ$), we obtain $v_{\alpha}(A)$ = 6.2925 mas yr$^{-1}$ and  $v_{\delta}(A)$ = $-$23.6355 mas yr$^{-1}$ ($v_{rot}$ = 24.4588 mas yr$^{-1}$ with PA = 165.09$^\circ$), and thus $\mu_{\alpha}(CM)$ = 570.4145 mas yr$^{-1}$ and $\mu_{\delta}(CM)$ = $-$36.3375 mas yr$^{-1}$ ($\mu(CM)$ = 571.5707 mas yr$^{-1}$ with PA = 93.65$^{\circ}$). We find that the estimated proper motions of the center of mass are very similar to those obtained with the full combined astrometric fit  $\mu_{\alpha}$ = 571.467 mas yr$^{-1}$ and  $\mu_{\delta}$ = $-$37.715 mas yr$^{-1}$ ($\mu$ = 572.72 mas yr$^{-1}$ with PA = 93.79$^\circ$; see Table~\ref{tab_3}). The difference in the estimated proper motions of the center of mass is about 1 mas yr$^{-1}$.
This further confirms that the full combined astrometric fit gives the best estimate for the proper motions of the center of mass of a binary system.

We can estimate the acceleration, $acc(A)$, of GJ~896A  due to the gravitational pull of the low-mass stellar companion GJ~896B, as follows:

\begin{equation}
acc(A) =  (2\pi/P_{AB})^{2} a_{A},
\end{equation}

\noindent
where $acc(A)$ is the mean acceleration of  GJ~896A, $P_{AB}$ is the orbital period of the binary system, and $a_{A}$ is the semi-major axis of the orbital motion of GJ~896A around the center of mass. Using the full combined solution (see column 3 of Table~\ref{tab_3}), we obtain that $acc(A)$  $\simeq$ 1.04234 mas yr$^{-2}$. In the case of the single-companion solution (see column 3 of Table~\ref{tab_2}), the fitted acceleration terms are $a_\alpha$ = 0.8873 mas yr$^{-2}$ and $a_\delta$ = 0.1402 mas yr$^{-2}$ , and thus the acceleration of the main star in the plane of the sky is $acc(A)$ = $\sqrt{a_\alpha ^{2} +  a_\delta^{2}}$ = 0.89831 mas yr$^{-2}$.
Thus, the estimated acceleration of GJ~896A due to GJ~896B is consistent with the acceleration found in the single-companion astrometric fit.

The acceleration $acc(A)$ obtained from the single-source astrometric fit of the VLBA data allows us to place a mass estimate for the  stellar companion, $m_{B}$, using

\begin{equation}
\left(\frac{m_{B}}{M_\odot}\right) = 0.02533  \left(\frac{acc(A)}{AU yr^{-2}}\right) \left(\frac{a_{AB}}{AU}\right)^{2}, 
\end{equation}

\noindent
or

\begin{equation}
\left(\frac{m_{B}}{M_\odot}\right) = 0.29368 \left[\left(\frac{acc(A)}{AU yr^{-2}}\right) \left(\frac{a_{A}}{AU}\right)^{2} \left(\frac{M_{AB}}{M_\odot}\right)^{2} \right]^{1/3},
\end{equation}

\noindent
where  $acc(A)$ = $\sqrt{a_{\alpha}^{2} + a_{\delta}^{2}}$ is the estimated acceleration needed to fit the absolute astrometric data (see column 3 of Table~\ref{tab_2}), $a_{AB}$ = $a_{A}$ + $a_{B}$ is the semi-major axis of the orbit of GJ~896B around GJ~896A, and m$_{AB}$ is the total mass of the binary system (see column 3 of Table~\ref{tab_3}). 
From equation~7 we find that the estimated mass of the stellar companion GJ~896B is $m_{B}$ $\approx$ 0.15728 M$\odot$. This estimated mass is consistent with the mass of the stellar companion ($m_{B}$ = 0.16527 M$\odot$) obtained from the full combined astrometric fit (see column 3 of Table~\ref{tab_3}).

\subsection{Expected Radial Velocities} \label{sec:radvel}

The solution of the full combined astrometric fit can be used to estimate an expected induced maximum RV of the star due to the gravitational pull of a companion as follows \citep[e.g.,][]{canto09,curiel20}:

\begin{equation}
K =\left(\frac{2 \pi G}{T}\right)^{1/3} \frac{m_{c} sin(i)} {(M_{*} + m_{c})^{2/3}} \frac{1} {\sqrt{1 - e^{2}}},
\end{equation}

\noindent
where G is the gravitational constant, and $T$, $M_{*}$, $m_{c}$, and $e$ are the estimated orbital period, star and companion masses, and the eccentricity of the orbit of the companion. Using the full combined astrometric solution (see column 3 of Table~\ref{tab_3}), 
the maximum RV of GJ~896A induced by the planetary companion GJ~896A$b$ is $K_{A}(b)$ $\sim$ 121 m s$^{-1}$, and the maximum RV induced by the stellar companion GJ~896B is $K_{A}(B)$ $\sim$ 867 m s$^{-1}$.
These RVs can in principle be observed with modern high-spectral resolution spectrographs. 

\begin{figure*}
\centering
\gridline{
              \fig{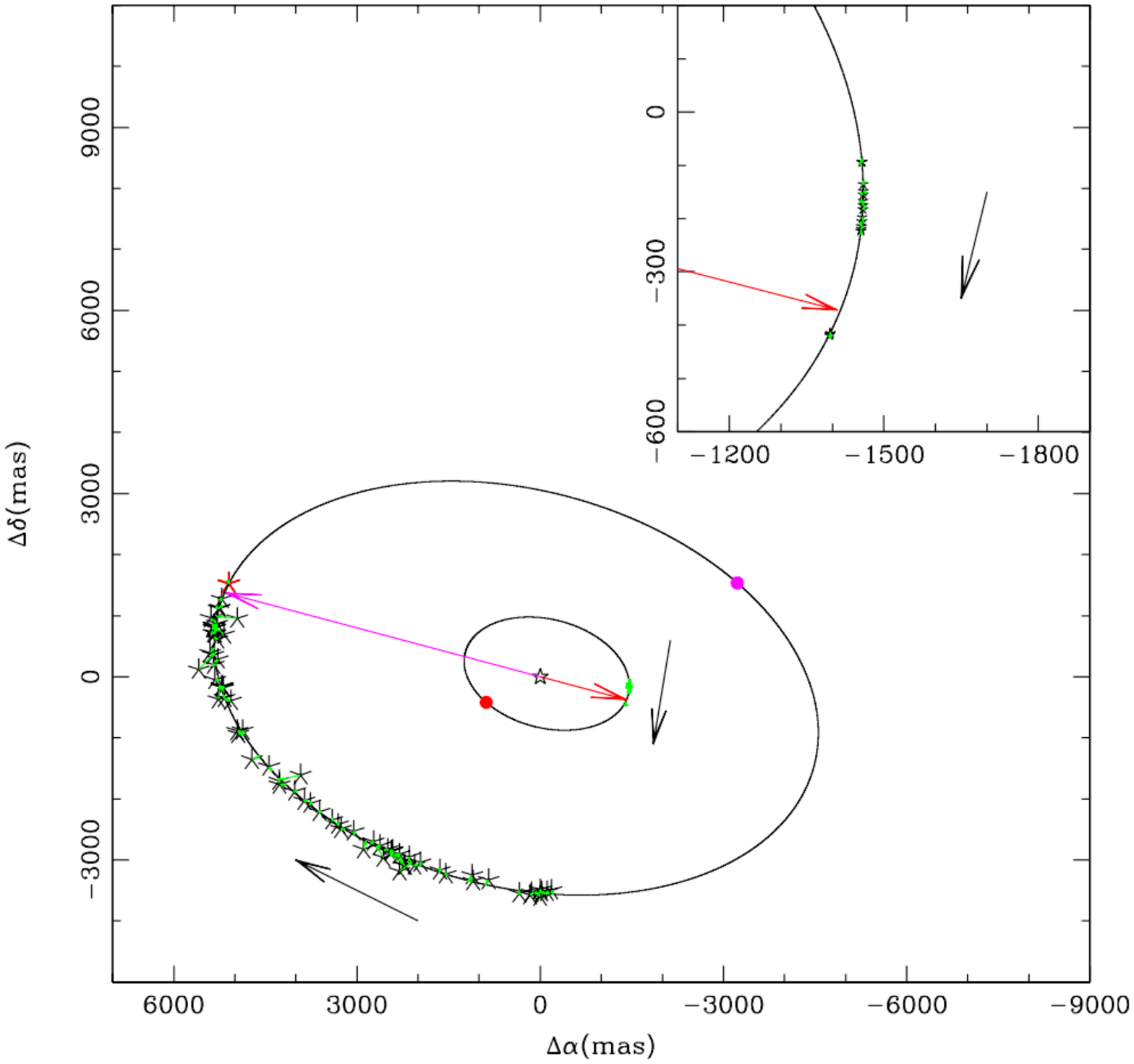}{0.48\textwidth}{a}
              \fig{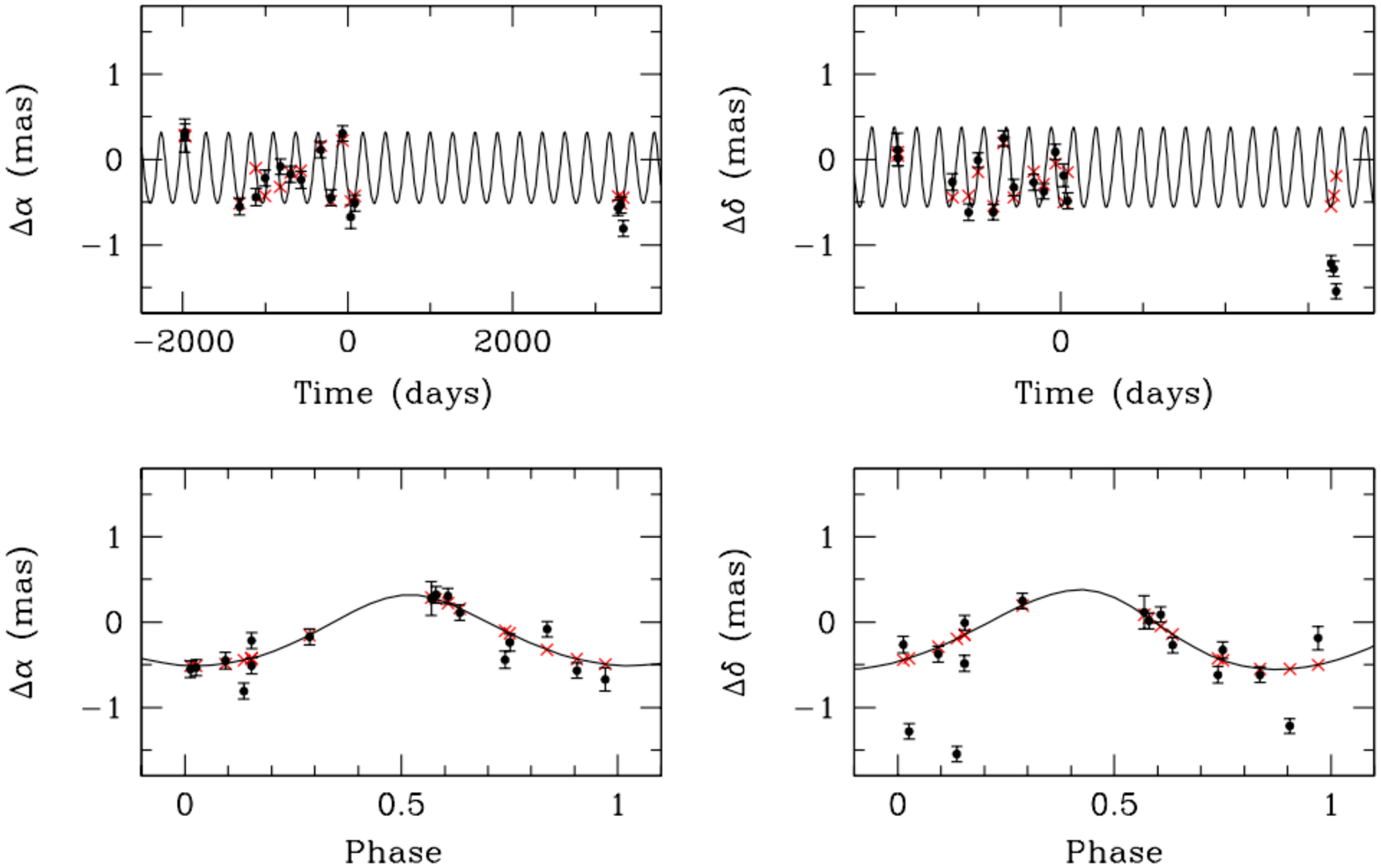}{0.48\textwidth}{b}
             }
\caption{Best fitted solution using the full combined astrometric fit of the binary system GJ~896AB. {\bf (a)}  Orbital motion of the stellar companion GJ~896B around the main star GJ~896A (big ellipse), and the orbital motion of the main star GJ~896A around the center of mass (inner ellipse). The black pointed stars show the observed optical/infrared position, and the two red pointed stars show the observed VLBA positions. The green lines connect the observed positions of the binary system with the expected position along the orbit at each observed epoch. The red and magenta dots indicate the position of the periastron of the orbits of the main star around the center of mass of the binary system and the orbit of the secondary star around the main star, respectively. The red and magenta arrows show the position of the ascending and descending nodes, respectively. The black arrows indicate the direction of the orbital motions. Note that the reference central point is different for both. In the case of the binary system, the central point indicates the position of the main star GJ~896A, while in the case of the orbital motion of the main star, the central position corresponds to the center of mass of the binary system. The inner panel shows an expanded section of the orbital motion of GJ~896A around the center of mass for the observed epochs. In this case, all the plotted epochs correspond to VLBA observations. {\bf (b)} Best fitted solution including the astrometric signal due to the planetary companion GJ~896A$b$, obtained from the full combined astrometric fit. The upper panels show the astrometric signal of the main star GJ~896A due to the planetary companion as function of time, with the parallax and proper motion subtracted, and taking into account the orbital motion of the stellar companion GJ~896B around the main star GJ~896A. The lower panels also show the astrometric signal of the main star GJ~896A due to the planetary companion, but in this case the signal is folded in phase. The black dots correspond to the absolute astrometric observations of the main star GJ~896A. The red crosses indicate the fitted position on the orbit for each observed epoch. The bars correspond to the observational astrometric errors.}
    \label{fig_8}%
    \end{figure*}

The maximum radial velocity of both stars occurred in October 2013, when GJ~896A and GJ~896B passed through the ascending and descending nodes, respectively, of their orbits around the center of mass of the binary system (see Figure~\ref{fig_8}).
The RV signal due to GJ~896A$b$ can be measured with modern high-spectral resolution spectrographs on a time span shorter than one year. 
Recent radial velocity observations of GJ~896A show a radial velocity variability of $\Delta V$ $\sim$ 175$\pm$37 m s$^{-1}$ \citep[e.g.,][]{gagne16}, which is consistent with the expected radial velocity of this source, induced by the planetary companion GJ~896A$b$. However, these observations were taken in just 7 epochs within a time span of about 4 years. Further observations will be needed to confirm whether the RV variability observed on GJ~896A is due to this planetary companion.

The RV signal of GJ~896A (and GJ~896B) due to the stellar companion will be difficult to measure due to the binary's very long orbital period of 229.06 yr. It will be very difficult to separate this signal from the systemic velocity $V_{0}$ of the binary system. However, we can use the observed mean RV of GJ~896A \citep[][]{lindegren18}, and the estimated maximum RV of this star due to its stellar companion to obtain a raw estimate of the systemic radial velocity of the binary system, as follows:

\begin{equation}
V_{obs}(A) = K_{A}(B) + V_{0}, 
\end{equation}

\noindent
where $V_{obs}$ is the observed radial velocity, $K_{A}(B)$ is the expected maximum RV of GJ~896A due to GJ~896B, and $V_{0}$ is the barycentric RV of the binary system. From the GAIA catalog, the observed radial velocity of GJ~896A is $V_{obs}(A)$ = $-$0.02$\pm$0.31 km s$^{-1}$  \citep[][]{lindegren18}. Since these RV observations of GJ~896A were take in a time span of about 2 years,  the RV signal from GJ~896A$b$ is averaged out in this mean radial velocity. The reference epoch of the GAIA DR2 observations is J2015.5, which is close to orbital location where the radial velocity of GJ~896A is maximum and negative (see Figure~\ref{fig_8}). 
Thus, $K_{A}(B)$  = $-$867 m s$^{-1}$, and the resulting barycentric RV of the binary system is $V_{0}$ $\sim$ 847 m s$^{-1}$.

A similar calculation can be carried out for  GJ~896B. In this case, the observed RV is $V_{obs}(B)$ = 3340 m s$^{-1}$  \citep[][]{morin08}, which was obtained also close to the maximum RV of this star. The reference epoch of the RV observations is 2006.0, which is close to orbital location where the radial velocity of GJ~896B is maximum and positive (see Figure~\ref{fig_8}). We estimate that the maximum RV of GJ~896B induced by the main star GJ~896A is $K_{B}(A)$ $\sim$ 2299 m s$^{-1}$. Thus, in this case we obtain a barycenter RV of $V_{0}$ $\sim$ 1041 m s$^{-1}$. This barycentric RV is different to that obtained from the RV observation of  GJ~896A (847 m s$^{-1}$). There is a difference of about 194 m s$^{-1}$. The RV measurements of GJ~896B were obtained in a time period of only a few days, and therefore, the observed RV would include the contribution due to possible companions associated to this M4.5 dwarf. 
This suggests that GJ~896B may have at least one companion. Such radial velocity signature should be easily detected using modern high-spectral resolution spectrographs. Recent observations also suggest that the M4.5 dwarf GJ~896B may be an unresolved binary system  \citep[][]{winters21}. Further observations will show whether this very low mass M dwarf star has a companion.

\subsection{Habitable zone and Snow Line}

A simple estimate of the habitable zone (HZ) can be obtained as follows. The inner and outer boundaries of the HZ around a star depends mainly on the stellar luminosity. Thus, combining the distance dependence of the HZ as function of the luminosity of the star and the mass-luminosity relation, the inner $a_{i}$ and outer $a_{o}$ radius of the HZ are:

\begin{equation}
a_{i} = \sqrt{ \left(\frac{L_\star}{L_\odot}\right)  \frac{ 1}{ 1.1}},
~~~~~ a_{o} = \sqrt{ \left(\frac{L_\star}{L_\odot}\right)  \frac{ 1}{ 0.53}}
\end{equation}

\noindent
\citep[e.g.,][]{selsis07, kopparapu13}, where $L_{\star}$ is the luminosity of the star in solar luminosities. In the case of M dwarfs with masses below 0.43 M$_\odot$, the mass-stellar luminosity relation can be approximated as:

\begin{equation}
\left(\frac{L_\star}{L_\odot}\right) =  0.23 \times \left(\frac{M_\star}{M_\odot}\right)^{2.3}
\end{equation}

\noindent
\citep[e.g.,][]{kuiper38, duric12}, where $M_{\star}$ is the mass of the star in solar masses. Using the estimated mass of GJ~896A ($m_{A}$ =  0.436 M$\odot$ ; see Table~\ref{tab_3}), the limits of the habitable zone around this M3.5 dwarf are $a_{i}$ $\sim$ 0.18 au and $a_{o}$ $\sim$ 0.26 au. This is considerably smaller than the semi-major axis of the orbit of the planetary companion GJ~896A$b$ ($a_{b}$ = 0.6397 au). Since the planetary companion is in an eccentric orbit ($e_{b}$ = 0.35), the minimum and maximum distance between the planet and the  star are 0.42 and 0.86 au, respectively. Thus, the orbit of the planet lyes outside the habitable zone of this M3.5 dwarf star.

We can also estimate if the planetary companion  GJ~896A$b$ is located inside the snow line, $a_{snow}$. The snow line is located at an approximate distance of \citep[e.g.,][]{kennedy08}

\begin{equation}
a_{snow} = 2.7 \left(\frac{M_\star}{M_\odot}\right)^{2}.
\end{equation}

\noindent
Using the estimated mass of GJ~896A, the snow line in this planetary system is located at a radius of $a_{snow}$ $\sim$ 0.51 au. Thus the estimated orbit of the planetary companion GJ~896A$b$ is located outside the snow line. 
However, given the eccentricity of the orbit, the planet moves around the estimated snow line distance, but most of the time is located outside the snow line.
Recent results suggest that stars with a mass of about 0.4 $M_{\odot}$, such as GJ~896A, have a 1\% probability of having at least one Jovian planet \citep[e.g.,][]{kennedy08}. Even when the probability of having a Jovian planet is very low, the results presented here show that the main star GJ~896A in the M-dwarf binary system GJ~896AB has at least one Jupiter-like planet.

\subsection{Flux variability of the source.}

\begin{figure*}
   \centering
    \plotone{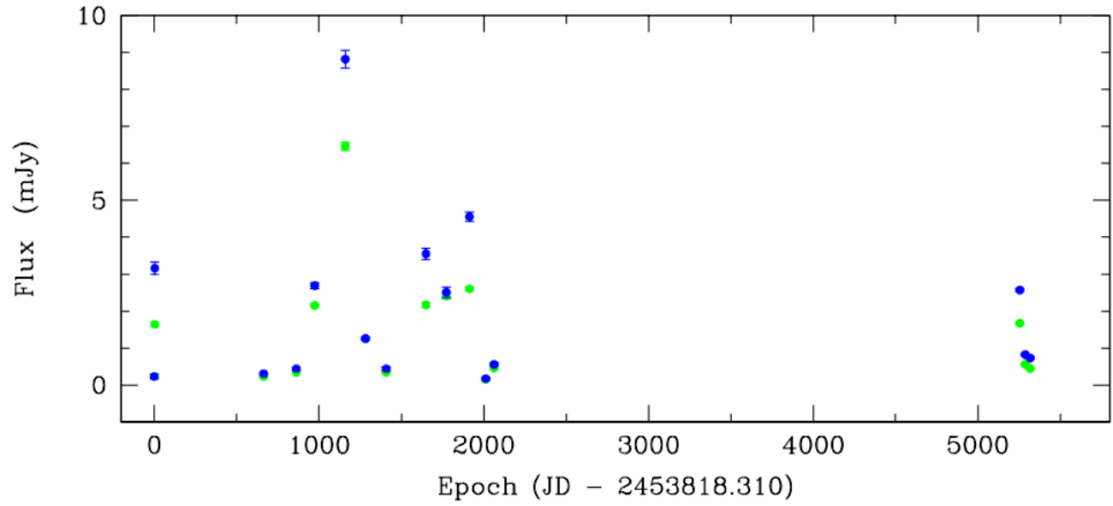}
    \caption{Radio flux density of GJ~896A as function of time. These VLBA observations were obtained at a frequency of 8.4 GHz. The green symbols correspond to the flux intensity of the source, and the blue symbols correspond to the integrated flux density of the source. The flux density observed on the source presents a large variability in short periods of time. There seem to be periods in time when the source is weak, having  a very low flux density ($\sim$ 0.2 mJy), and other periods of time when the source is active, having flux densities between 1 mJy and 9 mJy. We did not find a temporal pattern in the flux variability.}
    \label{fig_10}%
    \end{figure*}

The radio continuum flux density of GJ~896A is clearly variable in time. Figure~\ref{fig_10} shows that the source goes through a large variability in short periods of time. The flux variability does not seem to have a clear pattern. The flux density of the source has changed in nearly two orders of magnitud during the last 16 years of the radio observations, with variations at scales of months and at scales of a few years. However, further observations will be required to find if the variability has a defined temporal period.

\subsection{Mutual Inclination Angle} \label{sec:mutinc}

\begin{figure*}
\centering
\gridline{
              \fig{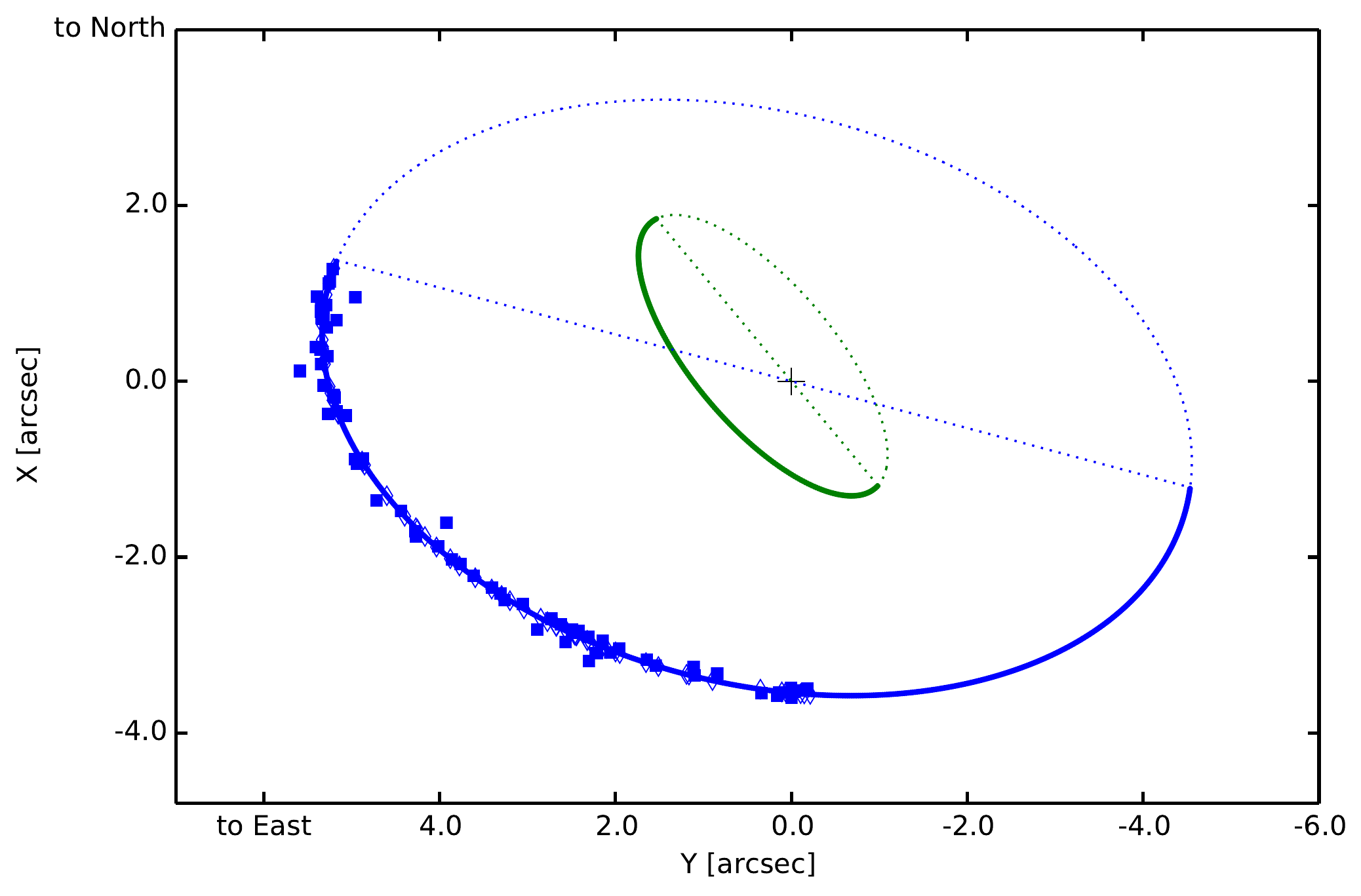}{0.48\textwidth}{a}
              \fig{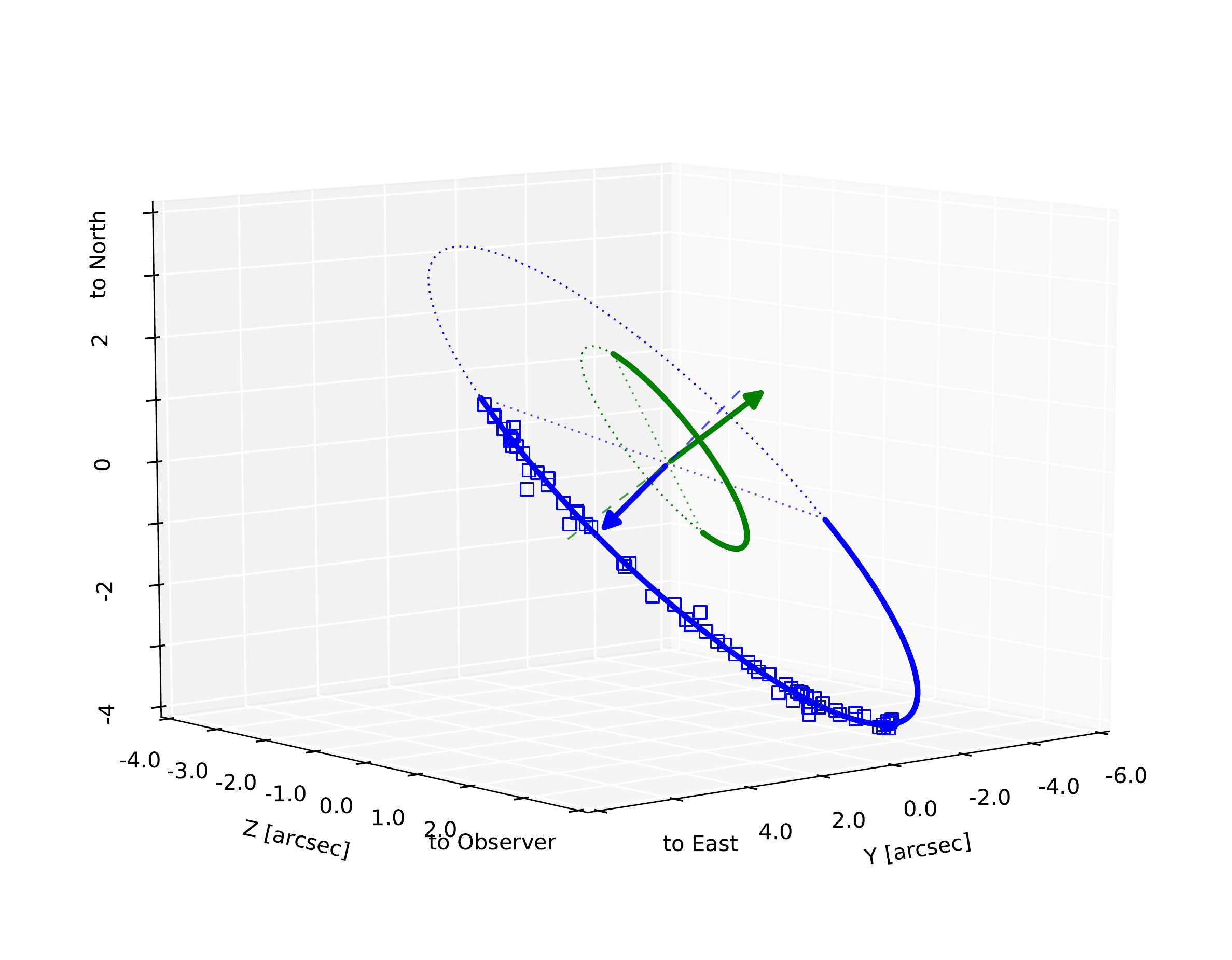}{0.48\textwidth}{b}
             }
\gridline{
              \fig{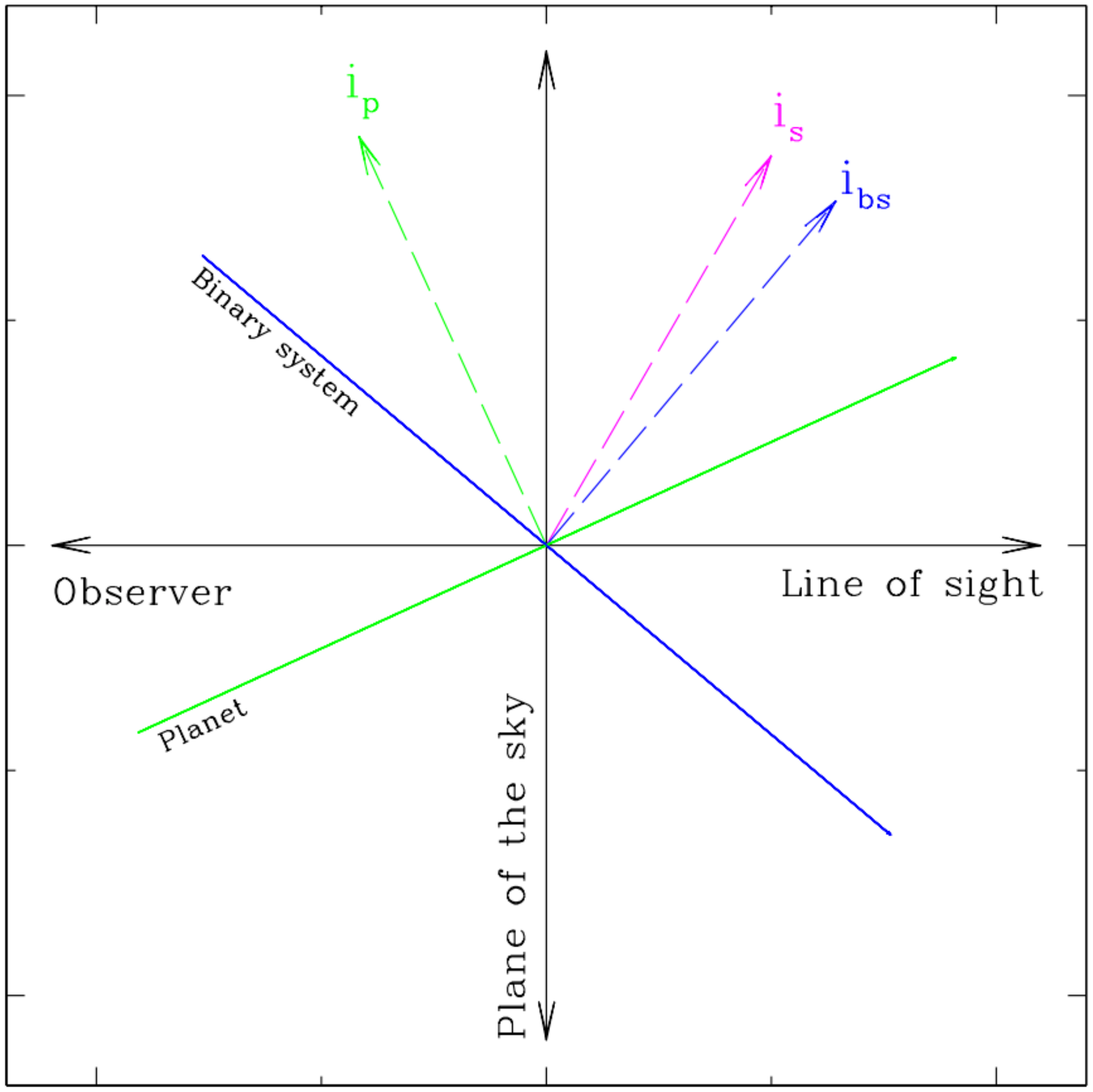}{0.48\textwidth}{c}
             }
\caption{3D orbital architecture of the binary system and planetary companion. {\bf (a)} 2-D plot of the fitted orbits of the binary system GJ~896AB (blue) and the planetary companion (green) projected on the plane of the sky. The orbit of the planet has been scaled by a factor of 20 to make easier the comparison. The blue squares show the observed relative position of GJ~896B around the main star GJ~896A. The thick lines indicate the side of the orbit that is closer to us, and the dotted lines indicate the side of the orbit that is on the other side of the plane of the sky. The straigth dotted lines show the line of nodes of both orbits. {\bf (b)} 3D plot of the fitted orbits of the binary system GJ~896AB (blue) and the planetary companion GJ~896A$b$ (green). The orbit of the planet has been scaled by a factor of 20 to make easier the comparison. The arrows indicate the direction of the rotation axis of both orbits. 
An animation presenting the 3D orbits is available at 
{\tt \url{https://drive.google.com/file/d/11ogBAzY_SdIHIeLPeHJi2PbEqf_3M__I/view?usp=sharing}}.
{\bf (c)} 2-D orientation (assuming $\Omega$ = 0) of the orbital planes (solid lines) and the direction of the rotation axis (discontinue arrows) of the orbital motion of the binary system (blue) and the planetary companion (green), projected on a plane formed by the line of sight (right),  the direction to the observed (left), and the East (down) and West (up) directions on the plane of the sky. The magenta arrow indicates the spin axis of the main star GJ~896A.}
\label{fig_9}%
\end{figure*}

Characterizing the full 3D orbital architecture of binary systems containing a planetary companion can aid to investigate the importance of the star-star and star-planet mutual interaction.
Combining relative and absolute astrometric data of the binary system, we are able to obtain the 3D orbital architecture of the system, including the planetary companion (see Figure~\ref{fig_9}).
Since the full combined astrometric fit (relative plus absolute astrometry) provides the inclination angle and the position angle of the line of nodes of the orbital planes of both the planetary companion GJ~896$b$ and the binary system GJ~896AB, we can directly measure the mutual inclination angle of this system. 
A first approach would be to estimate the inclination difference ($\Delta i$) between the inclination angles of the  orbital planes of the planet and the binary system, assuming that their position angle of the line of nodes is equal to 0$^{\circ}$. The inclination angles are measured with respect to the plane of the sky (such that $i$=0$^{\circ}$ corresponds to a face-on orbit, and it increases from the East toward the observer). Using the fitted inclination angles we obtain that $\Delta i =$ 60.9$^{\circ}$$\pm$22.5$^{\circ}$ (see Table~\ref{tab_3} and Figure~\ref{fig_9}), which is a very large difference. 
The mutual inclination angle $\Phi$ between the two orbital planes can be determined through 
\citep[e.g.,][]{kopal59, muterspaugh06}:

\begin{equation}
cos~\Phi = cos~i_{Ab} ~cos~i_{AB} +  sin~i_{Ab} ~sin~i_{AB} ~cos(\Omega_{Ab} - \Omega_{AB}),
\end{equation}

\noindent
where $i_{Ab}$ and $i_{AB}$ are the orbital inclination angles, and $\Omega_{Ab}$ and $\Omega_{AB}$ are the position angles of the line of nodes. The position angle of the line of nodes is measured anti-clockwise from the North toward the ascending node. 
Table~\ref{tab_3} contains the inclination angle and the position angle of the line of nodes for the planet GJ~896A$b$ and the binary system GJ~896AB. 
There is an ambiguity in the position angles of the line of nodes ($\Omega$ or $\Omega$ + 180$^{\circ}$, where $\Omega$  is the fitted angle), which can be disentangled by RV measurements. For the orbital motion of the binary system, recent RV measurements of both stellar components show that GJ~896B is receding ($V_{obs}(B)$ = $+$3.34 km s$^{-1}$;  \citet[][]{morin08}) and GJ~896A is approaching to us ($V_{obs}(A)$ = $-$0.02$\pm$0.31 km s$^{-1}$;  \citet[][]{lindegren18}), thus the correct position angle of the line of nodes is $\Omega_{AB}$ = 255.09$^{\circ}$ ($\Omega_{AB}$ + 180$^{\circ}$) (see Table~\ref{tab_3}). 
However, the planetary companion has no measured RV, so the fitted value of the position angle of the line of nodes of the orbital plane of GJ~896A$b$ could be either $\Omega_{Ab}$ = 45.6$^{\circ}$ or 225.6$^{\circ}$, the former is the fitted angle.
From these position angles we calculate a mutual inclination angle between the two orbital planes of $\Phi$ = 148$^\circ$  for the fitted $\Omega_{Ab}$, and  $\Phi$ = 67$^\circ$ for the second possibility ($\Omega_{Ab}$ + 180$^{\circ}$). 
Taking into account the long term stability of the system (see Sec.~\ref{sec:orbitstab}), we found that the former  solution  ($\Phi$ = 148$^\circ$) is stable in a very long period of time, while the latter solution ($\Phi$ = 67$^\circ$) is unstable in a short period of time.
This result indicates that there is a large mutual inclination angle of $\Phi$ = 148$^\circ$ between both orbital planes.

Recent observations suggest that the rotation axis of GJ~896A has an inclination angle of about 60$^{\circ}$$\pm$20$^{\circ}$ with respect to the line of sight \citep[][]{morin08}, and thus, an inclination angle of $i_{s}$ $\sim$ 210$^{\circ}$ with respect to the plane of the sky (see Figure~\ref{fig_9}). On the other hand, the inclination angle of the rotation axis of the orbital motion of GJ~896A$b$ is $i_{p}$ = 159.2$^{\circ}$$\pm$25.61$^{\circ}$ ($i_{b}$ + 90$^{\circ}$).
Thus these rotation planes have a difference in their inclination angles of $\Delta_{s-p}$ $\sim$ 51$^{\circ}$ (see Figure~\ref{fig_9}).
This indicates that the orbital motion of the planetary companion and the rotation plane of the star are far from being coplanar. 
We can also compare the angle of the rotation axis of the host star GJ~896A and the inclination angle, $i_{bs}$, of the rotation axis of orbital motion of the binary system GJ~896AB. In this case the inclination angle is $i_{bs}$ = 220.065$^{\circ}$$\pm$0.010$^{\circ}$ ($i_{AB}$ + 90$^{\circ}$). Thus the rotation planes of the star GJ~896A and the binary system have a difference in their inclination angles of $\Delta i_{bs-s}$ $\sim$ 10$^{\circ}$.
In addition, the rotation axis of GJ~896B, with respect to the line of sight, is also about 60$^{\circ}$$\pm$20$^{\circ}$ \citep[][]{morin08}, which is basically the same as that of GJ~896A, and thus, the difference between the inclination angle of the rotation axis of GJ~896B, with respect of the plane of the sky, compared with the rotation axis of the binary system is the same as that found for the star GJ~896A ($\sim$ 10$^{\circ}$). Thus, this suggests that the rotation planes of both stars are nearly parallel to the orbital plane of the binary system ($\Delta i$ $\sim$ 10$^{\circ}$), while the orbital planes of the planet and the binary system have a mutual inclination angle of $\Phi$ $\simeq$ 148$^{\circ}$. However, having similar inclination angles does not necessarily imply alignment between the rotation axis of both stars since the position angles of the line of nodes are unknown. The rotation axis of the stars could be different, and thus, they could have larger and different mutual inclination angles with respect to the orbital plane of the binary system.

\subsection{Orbital Stability} \label{sec:orbitstab}

We applied direct N-body integrations to the full combined orbital solution obtained from the Keplerian orbits of the binary system GJ~896AB and the planetary companion GJ~896A$b$ (see Table~\ref{tab_3}). We integrated the orbits for at least 100 Myrs using the hybrid integrator in Mercury~6  \citep[][]{chambers99}, which uses a mixed-variable symplectic integrator  \citep[][]{wisdom91} with a time step approximately equal to a hundredth ($\simeq 1/100$) of the Keplerian orbital period of the planetary companion. During close encounters, Mercury uses a Bulrich-Stoer integrator with an accuracy of $10^{-12}$. We identify an unstable system if: a) the two companions (the planetary companion and low-mass stellar companion) collide; b) the planet is accreted onto the star (astrocentric distance less than 0.003 au); and c) the planet is ejected from the system (astrocentric distance exceeds 200 au). 
The integration time of 100 Myrs long exceeds the 10,000 binary periods that is considered as a stability criterion.
The simulation using the fitted Keplerian orbital parameters proved to be stable for at least 100 Myrs. However, the simulation of the alternative position angle of the line of nodes of the planetary orbit ($\Omega = \Omega_{Ab} + 180^{\circ}$) turned out to be unstable in very short times, the planet  collided with the host star GJ~896A after a few tens of thousand years. In this case, the eccentricity of the planetary orbit  increases rapidly due to the interaction between the stellar companion and the planet. After a few tens of thousand years the eccentricity reaches an extreme value of 1, and thus, the planet collides against the main star. This indicates that the fitted solution contains the correct angle of the line of nodes of the planetary orbit. 

To complement our stability analysis, we also performed N-body long-term integrations using REBOUND  \citep[][]{rein12}. We tested the combined orbital solutions using two different integrators: IAS15  \citep[][]{rein15} and Mercurius  \citep[][]{rein19}. The first one is a 15th order high-precision non-symplectic integrator. The second one is a hybrid symplectic integrator. These two integrators allow us to corroborate the results obtained with Mercury 6. For both integrators, we integrate over 10,000 orbits of the binary system GJ~896AB using 20,000 points per orbital period (i.e., one sampling point every ~4 days). This allows us to monitor the changes in the orbital parameters of the planet GJ~896A$b$ with reliable accuracy. Both, the best-fit solution reported in Table~\ref{tab_3} and the complementary one with  $\Omega$ = $\Omega_{Ab}$ + 180$^{\circ}$ were tested. Our results are in agreement with those obtained with Mercury 6. Our best-fit solution is stable over the whole integration time, while the alternative solution becomes unstable in a short period of time.
A detailed discussion about the 3D orbital stability of this binary system is out of the scope of this paper and it will be presented elsewhere.

\subsection{Binary System Formation and Stability} \label{sec:binaryform}

The present separation between the two stars ($\sim$31.64 au) in this binary system suggests that they were most likely formed in a massive accretion disk by disk fragmentation, and not by turbulent fragmentation of the original molecular core. Binary systems formed by turbulent fragmentation are expected to have separations of hundreds, or even thousands, of au, which better explain the formation of binary systems with wide orbits \citep[e.g.,][]{offner16}. On the other hand, in the case of disk fragmentation, the lower mass stellar companion is formed in the outer parts of the disk (probably at a distance of a few hundred au) and then, during the early evolution of the binary system, the stellar companion migrates inwards to a closer orbit \citep[e.g.,][]{tobin16}. The apparently similar spin angles of the two stars ($\sim$ 210$^{\circ}$), and the relatively small, but significant, difference between the spin axis of both stars and the rotation axis of the binary system ($\ga$ 10$^{\circ}$), are consistent with this formation mechanism. 
Since we do not know the position angles of the line of nodes of the rotation plane of both stars, it is possible that the rotation planes are not coplanar, and thus, they probably have different mutual inclination angles with respect to the orbital plane of the binary system. 
Either way, the difference in the inclination angle between the spin axis of the stars and the orbital axis of the binary system ($\ga$ 10$^\circ$) suggests that during the evolution of this binary system, the star-star interaction between both stars may have significantly changed the orientation of the rotation axes of both stars. 
In addition, the large mutual inclination angle ($\Phi$ = 148$^\circ$) that we find between the orbital planes of the planetary companion GJ~896A$b$ and the binary system GJ~896AB indicates that something changed the orientation of the orbital plane of the planet, probably from initially being coplanar to be in a retrograde configuration at present time. The most likely origin of this large mutual inclination angle is the star-planet interaction, where the low mass stellar companion GJ~896B produces some important torque over the orbit of the planetary companion of GJ~896A. 

Recent studies have shown that in the case of planet-planet interaction (planetary systems with at least one external Jovian planet), the mutual inclination angles are $\la$10$^{\circ}$ \citep[see, e.g.,][]{laughlin02}. In a few cases there is evidence of an inclination between the orbits of the planets of up to $\sim$40$^{\circ}$ \citep[see, e.g.,][]{mcarthur10, dawson14, mills17}. However, since these studies are based on planetary systems detected with the RV and/or transit techniques, they lack of information about the position angle of the line of nodes, and thus, the mutual inclination angles calculated this way are lower limits. In addition, according to the catalogue of Exoplanets in Binary Systems \citep[][]{schwarz16},
160 planets have been found associated with 111 binary systems, 79 of which are in S-type orbits \citep[i.e., the planet is orbiting one of the stars; see, also][]{marzari19}. Most of these planets were found using RV and transit techniques, and some of the other planets were found using other techniques, such as imaging and microlensing. In all of them, some of the orbital parameters are missing, in particular, the position angle of the line of nodes of the orbital plane has not being obtained, and in several cases, the inclination angle is also unknown.
This is the first time that the remarkable full 3D orbital architecture of a binary-planetary system has been determined. 

\subsection{Planetary origin of the astrometric signal.}

The presence of a Jovian planet associated to the main star in a binary system, such as the one we present here, would produce secular perturbations on the lower mass stellar companion. Such interaction would produce long-term periodic variations in the orbital motion of the secondary (including the eccentricity and the inclination angle), as well as in the orbital parameters of the planetary companion. 
Given the orbital period  of about 229 yrs of this binary system, several decades of absolute and relative astrometric observations of the binary system are probably needed to fully constrain the orbital motion of the binary system.
As we have mentioned above, we are limited by the time span of the astrometric observations we present here (80 years for the relative astrometry and 16.6 yrs in the case of the absolute astrometry), which is much smaller than the orbital period of the binary system.
However, in the analysis we present here, we show that by combining the absolute astrometric observations of the primary and the secondary stars, as well as the relative astrometry of the binary system, in the astrometric fit, the orbital motion of the binary system and the planetary companion are well constrained.
We have found that the orbits of the binary system and the planetary companion are somewhat eccentric.
A similar astrometric signal due the planetary companion was found using different methods. 
We obtained a similar astrometric signal using both the periodogram of the astrometric data, and the single-companion astrometric fits of the absolute astrometric data obtained with two different algorithms (least-squares and AGA). This indicates that the astrometric signal is real. The detection of a similar astrometric signal with the full combined astrometric fit (relative plus absolute astrometry) further support the detection of the astrometric signal.
In addition, we complemented the analysis of the astrometric observations with the non-linear least squares minimization package {\tt lmfit} \citep[][]{newville20}, finding a similar astrometric solution (see Appendix).
This strongly indicates that the astrometric signal is real. Furthermore, the fact that the astrometric solution from the different methods is similar indicates that the astrometric signal is consistent with the companion being planetary in origin.
A detailed study of the stability of the orbital motions in this system can give important information about the star-star and star-planet  interactions. A detailed discussion about the 3D orbital stability of this binary system is out of the scope of this paper and it will be presented elsewhere.

The astrometric signal that we find is consistent with a planetary companion associated to the M3.5 Dwarf GJ~896A. However, this astrometric signal could be contaminated by the expected astrophysical $``$jitter$"$ added to the true source position due to stellar activity. It has been estimated that GJ~896A has a radius of $\sim$0.35 $R_\odot$  \citep[e.g.,][]{morin08, pearce20}. Thus, the radius of  GJ~896A at a distance of 6.2567 pc is about 0.260 mas. This result indicates that the astrometric signal due to the planetary companion (0.51 mas) is about twice the radius of the host star. In addition, assuming that the radio emission originates within $\sim$0.194 stellar radius above the photosphere \citep[e.g.,][]{liefke08, crosley18a}, the size of the expected $``$jitter$"$  is about 0.31 mas.
However, the analysis that we carried out for the variability of the radio emission of the main star GJ~896A shows that the $``$flares$"$ seen in several of the observed epochs contribute only with less that 0.12 mas to the expected $``$jitter$"$ (see Appendix~\ref{sec:flares} and panels (d) in Figure~\ref{fig_flares}), compared to the expected 0.31 mas contribution to the $``$jitter$"$ that we have estimated here. Thus, the expected $``$jitter$"$ due to the variability of the star is about a factor of 4.3 smaller than the astrometric signal of 0.51 mas observed in GJ~896A (see Table~\ref{tab_3}).
This result supports the detection of the planetary companion GJ~896A$b$. 

To further investigate the validity of the planetary astrometric signal, we have applied several statistical tests comparing the astrometric solutions obtained without and with a planetary companion (see Table~\ref{tab_2}). The solutions of the best astrometric fits show that when a planetary companion is not included the residuals of the fit have a rms scatter of 0.34 mas and a $\chi^2$ = 190.91, compared to a rms scatter of 0.20 mas and $\chi^2$ = 57.41 when the planetary companion is included in the fit.  An  F-test shows that the probability of the $\chi^2$ dropping that much (due to the inclusion of the planetary companion) is less than 4\% by mere fluctuations of noise. Using the Bayesian Information Criterion (BIC), we find that the inclusion of the planetary companion is preferred with a $\Delta$BIC = $-$109.24, this is a significant difference between our best fit model with the planetary companion and the one without it. Statistically, an absolute value of Delta BIC of more than ten implies that the model with the planetary companion better reproduces the data without overfitting it by including more free parameters in the model. Similar results were obtained using the Akaike Information Criterion (AIC) with a $\Delta$AIC = $-$119.50. We therefore conclude that there is a very high probability that the planetary companion GJ~896Ab is orbiting the main star GJ~896A.

The large proper motions of the binary system may also contribute to the expected $``$jitter$"$ of the star. The change in position of GJ~896A due to the proper motions of the binary system, $\Delta_{PM}(A)$ in mas, can be estimated by:

\begin{equation}
\Delta_{PM}(A) =  \sqrt{\mu_\alpha^{2} + \mu_\delta^{2}} \times \left(\delta t / 8766\right),
\end{equation}

\noindent
where $\delta t$ is the time span of each observed epoch in hours (between 3 and 5 hours on source). Using the estimated proper motions of the binary system ($\mu$ = 572.71 mas yr$^{-1}$), we obtain that the maximum expected contribution to the $``$jitter$"$ by the proper motions of the system is $\Delta_{PM}(A)$ = 0.16 mas in a time span of 2.5 hours (half the observing time), which is about three times smaller than the observed astrometric signal. We also obtain that the contribution of the orbital motion of GJ~896A around the center of mass to the expected $``$jitter$"$ of the star (about 0.009 mas in a time span of 2.5 hours) is smaller than that estimated from the proper motions of the star. Adding in quadrature all these possible contributions, the total expected $``$jitter$"$ is about 0.2 mas, which is still about a factor of 2.6 smaller than the astrometric signal due to the planetary companion GJ~896A$b$. The highest contribution to the expected $``$jitter$"$ is that due to the proper motions of the star.  
However, the astrometric signal is observed in both direction (R.A. and Dec.), while the proper motion of the star is basically in the east direction. In addition, the contribution of the proper motions of the star to the expected $``$jitter$"$ probably averages out when we integrate over the time span of the observations. In addition, given the synthesized beam of the images (about 2.7$\times$1.1 mas, in average), the magnitude of the proper motions is in fact too small to even affect the estimated size of the source. This suggests that the contribution of the proper motions to the expected $``$jitter$"$ is probably much smaller than we have estimated here. Thus, the expected $``$jitter$"$  that we estimate here is most likely an upper limit. 
This  further indicates that the astrometric signal is real, and planetary in origin.

\section{Conclusions and Final Remarks}  \label{sec:conclusions}

The combined (relative plus absolute) fit of the astrometric observations of this binary system show that the main star GJ~896A has at least one planetary companion. 
This is the first time that a planetary companion of one of the stars in a binary system has been found using the astrometry technique.
The astrometric solution indicates that the planetary companion has an orbital period of 284.39 $\pm$ 1.47 days, an estimated mass of 2.26 $\pm$ 0.57 M$_{J}$, a relatively eccentric orbit ($e_{Ab}$ = 0.35 $\pm$ 0.19), and a semi-major axis of $a_{b}$ = 0.63965 $\pm$ 0.00067 au (or 102.27 $\pm$ 0.15 mas). In addition, the full combined astrometric fit also shows that the binary system has an estimated orbital period of 83664.63 $\pm$ 1.98 days (or 229.06 yrs), and the two stars have estimated masses of $m_{AB}$ = 0.603410 $\pm$ 0.000025 $M_{\odot}$, 
$m_{AB}(A)$ = 0.43814 $\pm$ 0.00065 $M_{\odot}$ 
and $m_{AB}(B)$ = 0.16527 $\pm$ 0.00025 $M_{\odot}$, respectively. The astrometric solution also indicates that the binary system and the stars have semi-major axis of $a_{AB}$ = 31.635 $\pm$ 0.033 au, $a_{AB}(A)$ = 8.6646 $\pm$ 0.0091 au and $a_{AB}(B)$ = 22.971 $\pm$ 0.024 au, respectively.
Thanks to the proximity of the binary system GJ~896AB, the Jovian-like planetary companion GJ~896A$b$ $-$ one of the nearest to Earth yet found $-$ is well suited for a detailed characterization (for example, direct imaging and spectroscopy) that could give important constrains on the nature and formation mechanisms of planetary companions in close binary systems.

Combining the relative and absolute astrometric observations, we have found the 3D orbital architecture of the binary system GJ~896AB and the planetary companion GJ~896A$b$. 
We have performed long-term numerical integrations to test the stability of the orbital solution of this binary system, using both posible position angles of the line of nodes of the planetary companion. We find that only one solution is stable. The second solution of the Keplerian orbit, using $\Omega$ = $\Omega_{Ab}$ + 180$^{\circ}$, turns out to be unstable in very short timescales, the planetary companion collides with the host star GJ~896A after a few thousand years.  
On the other hand, the fitted solution ($\Omega = \Omega_{Ab}$) proved stable for at least 100 Myrs. This indicates that the position angle of the line of nodes of both orbital planes are $\Omega_{AB}$ = 255.1$^{\circ}$ and $\Omega_{Ab}$ = 45.6$^{\circ}$, and their mutual inclination angle is $\Phi$ = 148$^\circ$. 
This result is consistent with both orbits being retrograde.
This is the first time that the full 3D orbital architecture of a binary system with a planetary companion has been obtained using astrometric observations. This kind of results can not be achieved with other exoplanet methods. Astrometry gives important complementary information to other exoplanet detection techniques. In addition, high-angular resolution radio astrometry is becoming a powerful technique, capable of giving the full 3D orbital architecture of  planetary systems and planetary systems in binary and multiple stellar system.

Our results demonstrate that astrometric observations have the potential to fully characterize the orbital motions of individual and multiple planetary systems, as well as the 3D orbital architecture of binary systems, and binary systems with planets associated to them.
The discovery of gaseous planets associated to low-mass stars poses a great challenge to all current planetary formation scenarios. For instance, core accretion models face serious problems to explain giant planets orbiting around M dwarfs with masses below 0.4 M$_\odot$  \citep[e.j.,][]{laughlin04,ida13,burn21}. In order to explain such planets, these models need to include extraordinary conditions, such as increasing the efficiency of core-accretion planet formation, using high mass protoplanetary disks, which are inconsistent with observational results, and/or slowing down (reducing) their migration speed. 
It is not clear if gravitational instability of the protoplanetary disk \citep[e.j.,][]{mercer20,boss21} could more naturally produce giant planets around low-mass stars. 
The formation of this kind of planetary systems through disk fragmentation also requires high-mass protoplanetary disks (with high accretion rate from an envelope).
Furthermore, the discovery of Jovian planets associated with low-mass binary systems, such as the one we have found, is even more challenging to current formation scenarios. Particularly, in the case of close binary systems with separation $<$40 au, where it is expected that the stellar companion would truncate the protoplanetary disk, Jovian planets will be very difficult to form.
Further theoretical models will be required to understand the formation of giant-mass planets, such as the one we found associated to the main star of the low-mass binary system GJ~896AB.
In addition, since most stars are in binary or multiple systems, our understanding of systems such as this one will be crucial to understand the phenomenon of planetary formation in general.


~~
~~
~~

\begin{acknowledgements}
\noindent
We are grateful to the anonymous referee for the useful comments and suggestions, which helped improve this paper.
We thank L. F. Rodr{\'\i}guez for valuable comments on an early version of the paper.
S.C. acknowledges support from UNAM, and CONACyT, M\'exico. 
The authors acknowledge support from the UNAM-PAPIIT IN103318 and IN104521 grants.
The observations were carried out with the Very Long Baseline Array (VLBA), which is part of the National Radio Astronomy Observatory (NRAO). The NRAO is a facility of the National Science 
Foundation operated under cooperative agreement by Associated Universities, Inc.
This research has made use of the Washington Double Star Catalog maintained at the U.S. Naval Observatory.
This research has made use of the Catalogue of Exoplanets in Binary Systems maintained by Richard Schwarz and \'A. Bazs\'o at {https://www.univie.ac.at/adg/schwarz/bincat$_{-}$binary.html}. 
This publication makes use of the SIMBAD database operated at the CDS, Strasbourg, France.
This work has made use of data from the European Space Agency (ESA) mission Gaia ({https://www.cosmos.esa.int/gaia}), processed by the Gaia Data Processing and Analysis Consortium (DPAC, {https://www.cosmos.esa.int/web/gaia/dpac/consortium}). Funding for the DPAC has been provided by national institutions, in particular the institutions participating in the Gaia Multilateral Agreement.
\end{acknowledgements}

\vspace{5mm}
\facilities{VLBA}

\software{AIPS \citep[][]{greisen03}, 
               astropy \citep{astropycol13,astropycol18},
               corner\citep{foremanmackey16}, 
               emcee \citep{foremanmackey13}, 
               lmfit \citep[][]{newville20}, 
               scipy \citep[][]{virtanen20},
               matplotlib,\citep{hunter07},
               numpy\citep{vanderwalt11}. 
               PyAstronomy\citep{czesla19}.
          }

~~
~~
~~

{~~~~~~~~~~~~~~~~~~~~~{\bf ORCID iDs}}

Salvador Curiel {https:/orcid.org/0000-0003-4576-0436}

Gisela N. Ortiz-Le\'on {https:/orcid.org/0000-0002-2863-676X}

Amy J. Mioduszewski {https:/orcid.org/0000-0002-2564-3104}

Joel Sanchez-Bermudez {https:/orcid.org/0000-0002-9723-0421}

~~
~~
~~

\appendix

\section{Variability of the radio emission}\label{sec:flares}

\begin{figure}[t]
\begin{center}
  \plotone{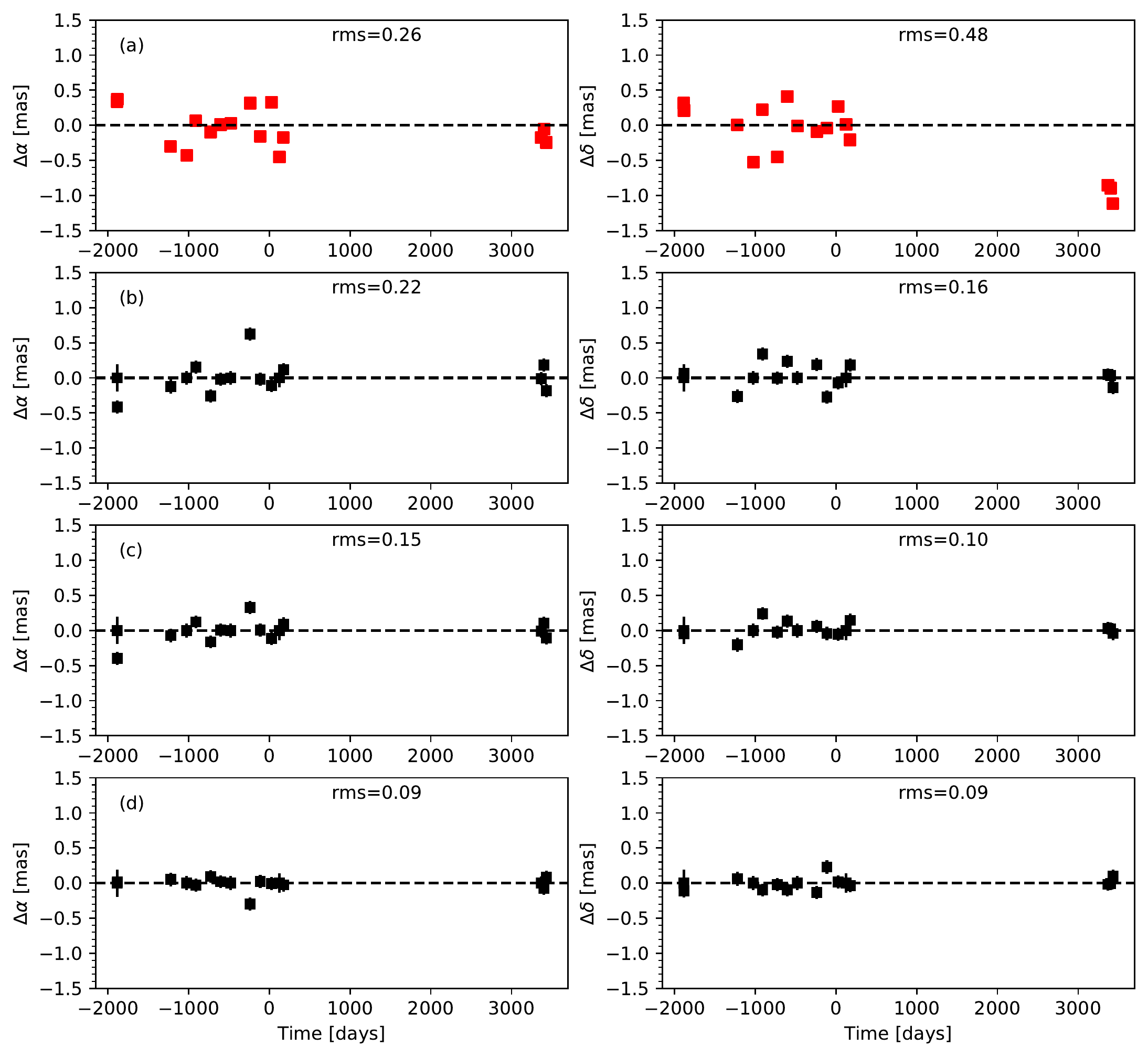}
\end{center}
  \caption{(a) Residuals obtained by subtracting parallax, proper motions and orbital motion of the stellar companion from the GJ896A positions using the combined astrometric fit. (b) Offsets between the position of the flare and the source position without the flare (the quiescent emission). (c) Offsets between the position of the flare and the average source position. (d) Offsets between the source position without the flare (the quiescent emission) and the average source position. The rms in each case is indicated at the top of the sub-panels.}
  \label{fig_flares}%
\end{figure}

In this section, we investigate the level of intra-epoch variability of the radio emission detected with the VLBA and how it affects the astrometry of GJ~896A. To that end, the real part of the $(u,v)$ data was plotted as a function of time, averaged over the $(u,v)$ plane. At twelve epochs, we found short-timescale ($\sim$7--40 min) variability or flares in the light curves of the real part of the visibility. During these flares, the peak flux density of the source increases by about 1.2 to 6 times above the mean flux density. 
From these flux curves we determine the time interval where the flare is observed, and construct maps using only the time range of the flare. Using these images, we then obtain the position of the flare with the AIPS task {\tt MAXFIT}. We also construct maps using only the time range outside the flare (which we will call ``quiescent'' emission) and measure the source position. 
Panel (a) in Figure \ref{fig_flares} shows residuals obtained for GJ~896A after removing  parallax, proper motion and orbital motion due to the stellar companion GJ~896B from the full combined astrometric fit solution (see Section~\ref{sec:caf} and Figure \ref{fig_7}a). Here, it is important to note that the positions of GJ~896A reported in Table~\ref{tab_1} were obtained using the maps constructed for the full observing time at each epoch (typically 3--5 hours), which we will call the ``average'' source position.  
In the next three panels of Figure \ref{fig_flares}, we plot offsets between (b) the position of the flare and the source position without the flare, (c) the position of the flare and the average source position, and (d) the source position without the flare and the average source position. We see in panel (b) that at some epochs there are big offsets between the flare and the source position without the flare, i.e., the quiescent emission. These offsets have rms values of 0.22 and 0.16~mas, in R.A.\ and Dec., respectively. The plots in panel (d), on the other hand, suggest that when we average the data over the full observing time, the ``average'' position is dominated by the quiescent emission and not by the flare. In this case, the differences between the position of the quiescent emission and the average position are small (with rms values of 0.09~mas in both R.A.\ and Dec.) compared to the offsets observed during the flare events (see panel (b)). In addition, the differences plotted in panel (d) are also small compared to the residuals from the astrometric fit shown in panel (a), which correspond to the astrometric signal of the main star due to the planet. The rms of the differences plotted in panel (d) are 2.9 and 5.8 times smaller,  in R.A.\ and Dec., respectively, than the rms of the residuals. We also note that these differences do not follow the same temporal trend as the residuals, which indicate that the residuals of the astrometric fit of our data cannot be induced by the flare activity. From this analysis, we conclude that the flares occurring at short-timescales do not affect the astrometry of GJ~896A. This is because the position of the radio emission obtained by averaging over the full observing time of each observation is dominated by quiescent emission. Furthermore, this indicates that the astrometric residuals have a different origin, such as the presence of one or more companions.

The flare events observed in GJ~896A  are short$-$ and long$-$duration bursts with time scales between 7 and 40 minutes. The radio emission shows right-circular polarization (RCP) during the outbursts, with a degree of polarization between 10$\%$  and 80$\%$, except one epoch where the radio emission shows left-circular polarization (LCP). These characteristics suggest that the origen of the busts may be associated to electron cyclotron maser emission (ECM). This tentative interpretation is consistent with the interpretation of the outbursts events observed in this magnetically active M dwarf star with the Jansky Very Large Array \citep[e.g.,][]{crosley18a, crosley18b, villadsen19}. However, there may also be multiple phenomena responsible for the short-duration bursts that we observed in GJ~896A. A detailed discussion of the possible origin of the observed radio flares is out of the scope of this paper and it will be presented elsewhere.

\section{Posterior Sampling} \label{sec:mcmc}

\begin{figure*}
\centering
\gridline{
              \fig{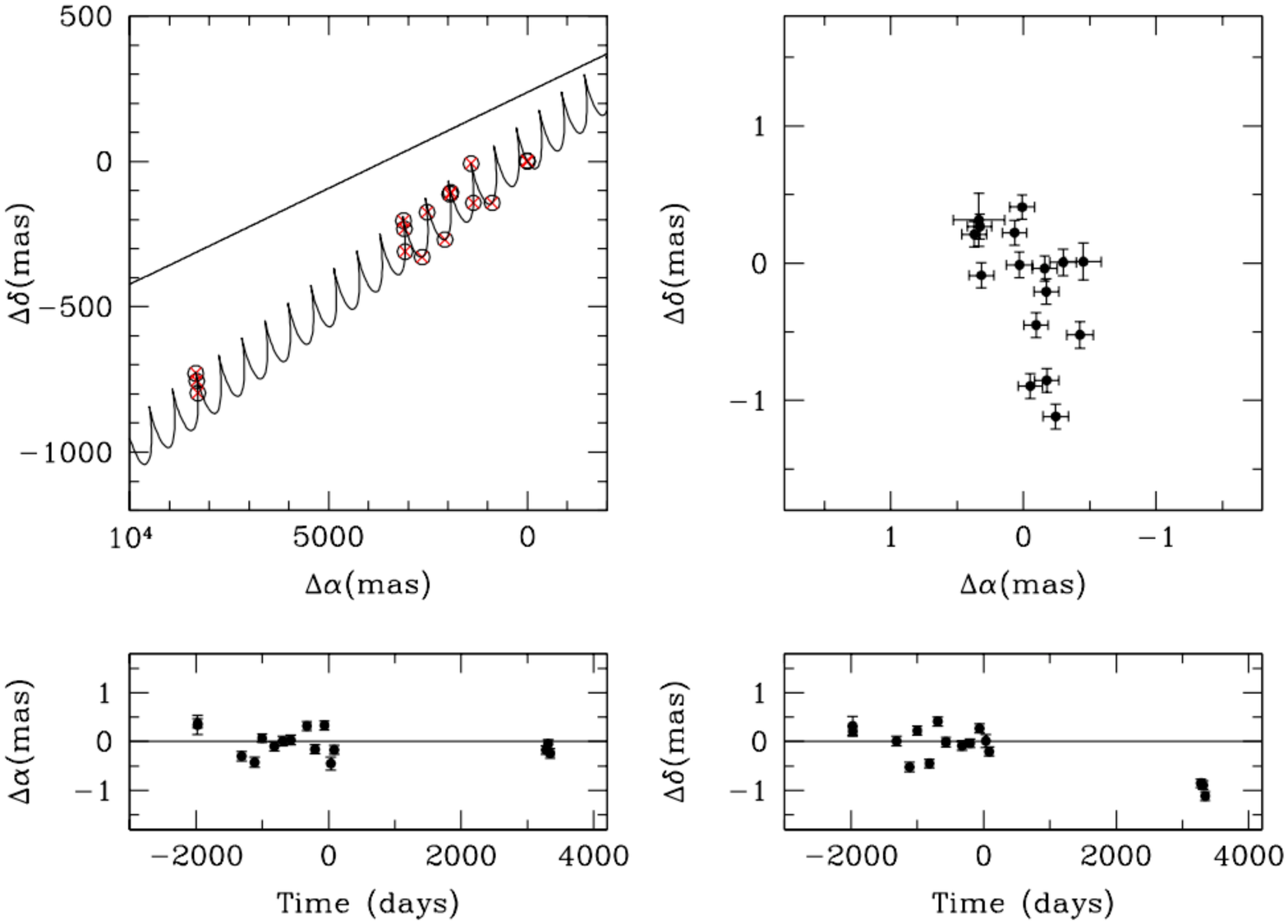}{0.48\textwidth}{a}
              \fig{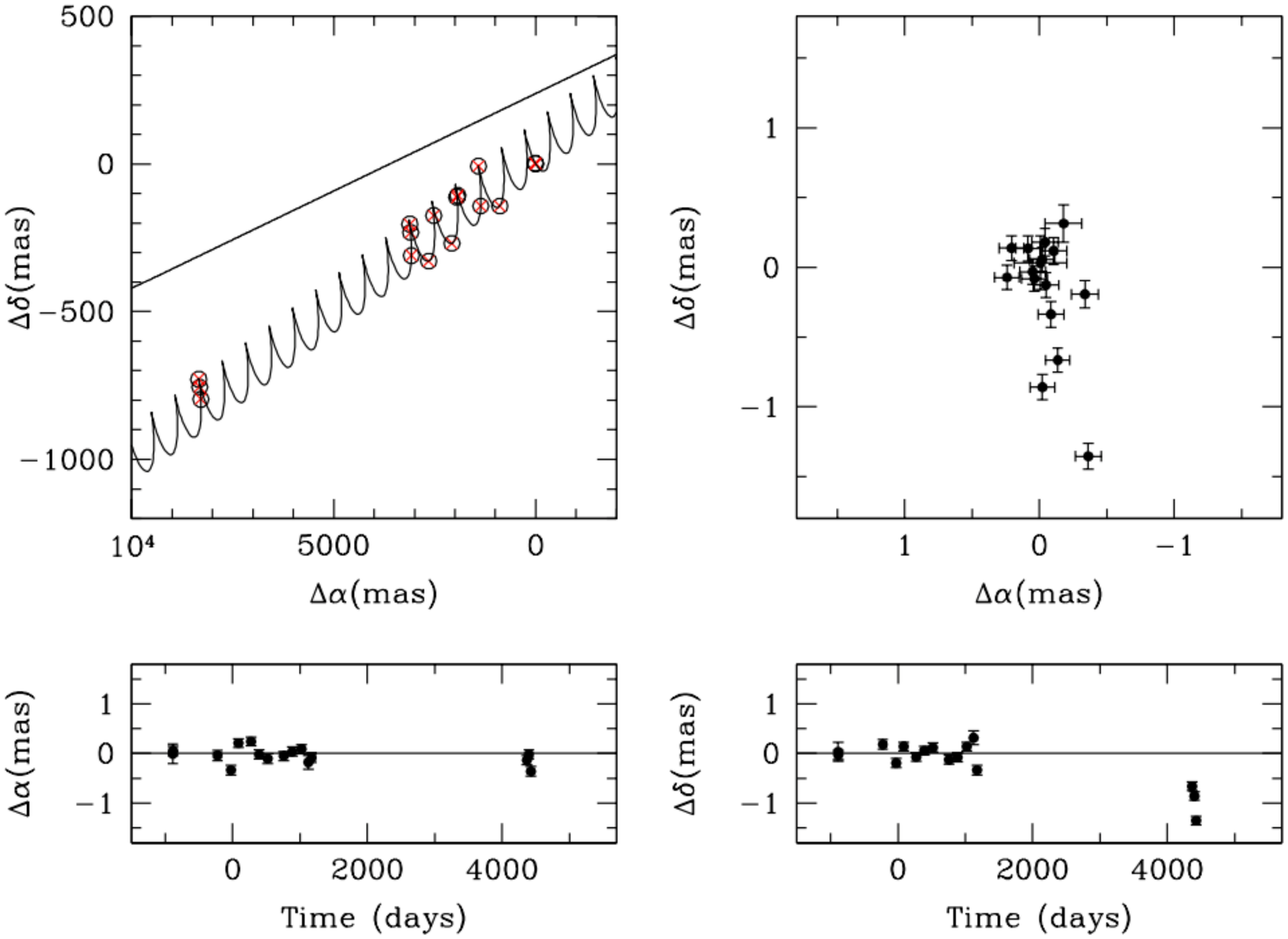}{0.48\textwidth}{b}
             }
\caption{These plots show the combined (relative plus absolute) astrometric fit of the binary system GJ~896AB. {\bf (a)}  Astrometric fit of the main star GJ~896A, obtained from the combined astrometric fit of the binary system GJ~896AB using both, the optical and infrared observations for the relative astrometry, and the VLBA observations for the absolute astrometry of both stars GJ~896A and GJ~896B. The upper-left panels show the observed data and the astrometric fit. The stright line in the upper-left panel corresponds to the direction of the proper motions of this binary system. The right and lower panels show the residuals in R.A. and Dec. as function of time. The residuals show a clear short-term trend that suggests that it could be due to at least one companion. The residuals also seem to suggest a long-term trend that could be due to a planetary companion with a large orbital period. {\bf  (b)} Similar to the left figure, but in this case the panels show the astrometric fit of the main star GJ~896A, obtained with the full combined astrometric fit of the astrometric data of the binary system GJ~896AB. In this case the residuals are smaller than those shown in the left figure. Notice that the rms and the distribution of the residuals shown in {\bf (a)} and {\bf(b)} are very similar to those shown in Fig.~\ref{fig_5}.}
    \label{fig_7}%
    \end{figure*}

\begin{figure*}
   \centering
    \plotone{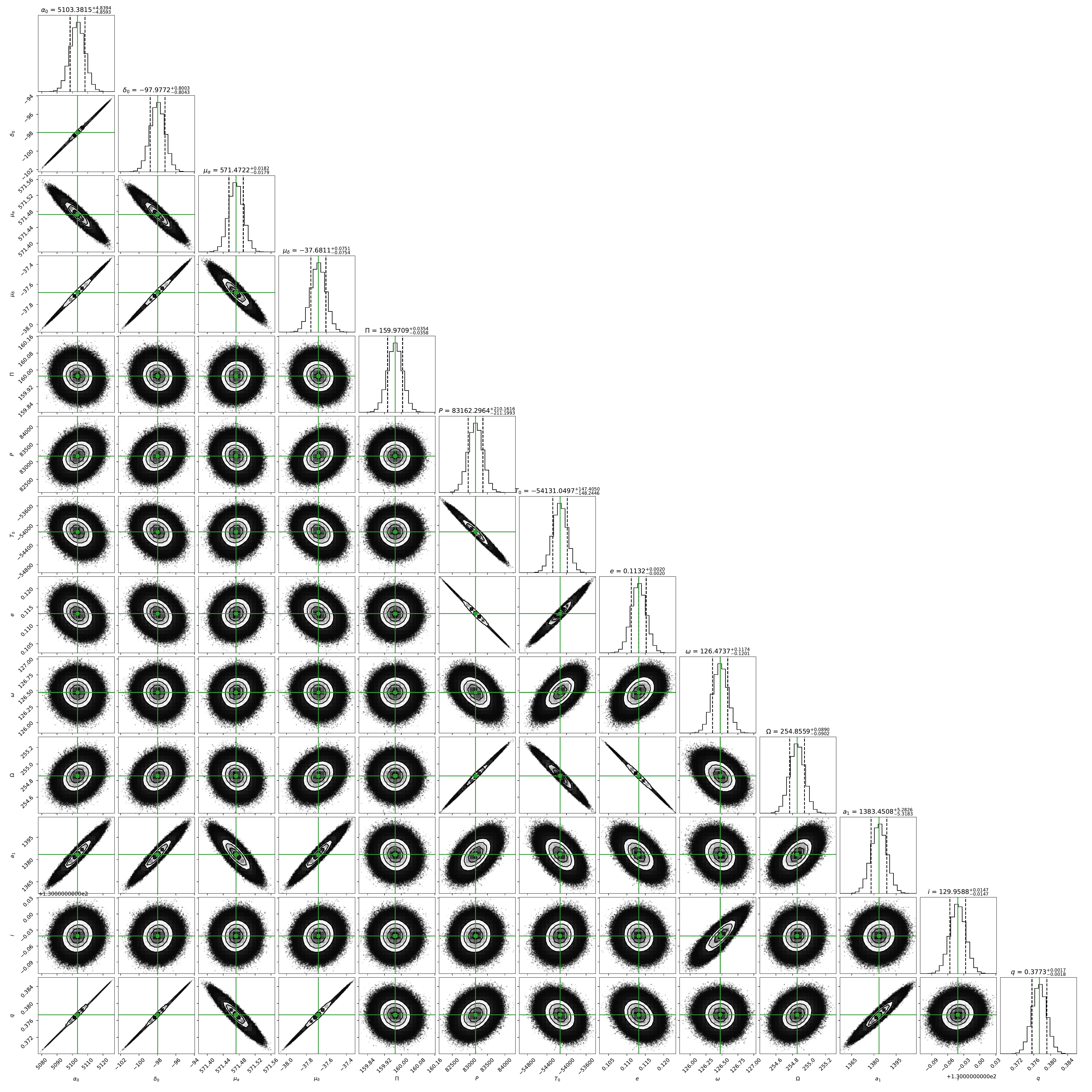}
    \caption{Correlations between the fitted parameters from the MCMC analysis using the corner code. On top of each column, we show the 2D posterior probability histogram of each fitted parameter. The green lines indicate the mean value of each fitted parameter. The doted lines indicate the $\pm1\sigma$ estimated errors.}
    \label{fig_emcee}%
    \end{figure*}

\begin{figure*}
   \centering
    \plotone{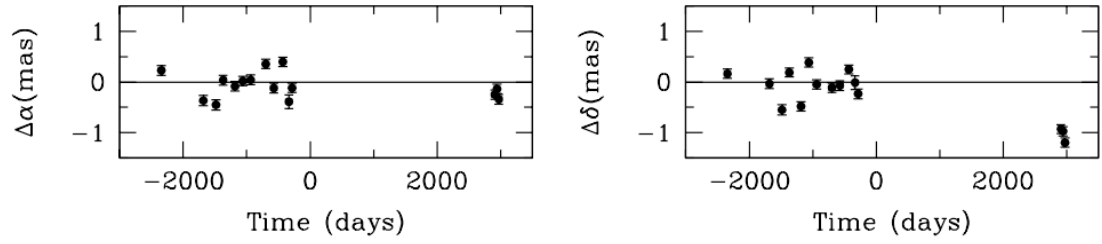}
    \caption{Residuals from the combined astrometric fit using {\tt lmfit}.}
    \label{fig_emcee_resid}%
    \end{figure*}

We used the open-source package $lmfit$ \citep[][]{newville20}, which includes several minimization algorithms to search for the best fit of observational data. In particular, we used the default Levenberg-Marquardt minimization algorithm, which uses a non-linear least-squares minimization method to fit the data. This gives an initial fit solution to the astrometric bidimensional data.
{\tt lmfit} also includes a wrapper for the Markov Chain Monte Carlo (MCMC) package {\tt emcee} \citep[][]{foremanmackey13}. When fitting the combined astrometric data (absolute astrometry of both stars GJ~896A and GJ~896B and the relative astrometry of the binary system GJ~896AB), we weighted the data by the positional errors of both coordinates ($\alpha$ and $\delta$). We used 250 walkers and run the MCMC for 30000 steps with a 1000 step burn-in, at which point the chain length is over 50 times the integrated autocorrelation time. The fitted solution is listed in Table~\ref{tab_A1}, and Figure~\ref{fig_emcee} shows the correlation between the fitted parameters. The fitted solution is very similar to that obtained from the full combined astrometric fit (see column (2) in Table~\ref{tab_3}). The $\chi^2$ and reduced $\chi^2$ are also very similar to those obtained from the  full combined astrometric fit. The residuals of the fit are shown in  Figure~\ref{fig_emcee_resid}, which are very similar to those obtained with the non-linear least-squares and AGA algorithms (see Figure~\ref{fig_7}a).

{}




\begin{deluxetable}{cccccccc}
\centering                          
\tablecaption{Properties of the VLBA detections. \label{tab_1} }
\tablewidth{0pt}
\tablehead{
\colhead{Julian day} & \colhead{$\alpha(J2000)$} & \colhead{$\sigma_\alpha$} & \colhead{$\delta(J2000)$} 
&  \colhead{$\sigma_\delta$}  & \colhead{Flux density} & \colhead{rms}  \\
& \colhead{(h:m:s)}& \colhead{(s)} & \colhead{($^{\rm o}$:$'$:$''$)} & \colhead{($''$)} & \colhead{(mJy)} 
& \colhead{($\mu$Jy~beam$^{-1}$)} \\
\colhead{(1)}  & \colhead{(2)}  & \colhead{(3)}  & \colhead{(4)}  & \colhead{(5)}  & \colhead{(6)}  & \colhead{(7)} 
}  
\startdata
\sidehead{GJ~896A}
\hline
2453818.30167 & 23:31:52.4337116 & 0.0000130 & +19:56:13.712431 & 0.000193 & 0.23 $\pm$ 0.06 &  37 \\
2453821.29347 & 23:31:52.4345726 & 0.0000063 & +19:56:13.715392 & 0.000092 & 3.16 $\pm$ 0.16 &  63 \\
2454482.38363 & 23:31:52.4967295 & 0.0000066 & +19:56:13.570044 & 0.000097 & 0.31 $\pm$ 0.02 &  11 \\
2454679.86273 & 23:31:52.5343755 & 0.0000067 & +19:56:13.704862 & 0.000098 & 0.44 $\pm$ 0.04 &  18 \\
2454792.59143 & 23:31:52.5301869 & 0.0000061 & +19:56:13.570540 & 0.000089 & 2.69 $\pm$ 0.08 &  40 \\
2454978.01521 & 23:31:52.5703946 & 0.0000061 & +19:56:13.606475 & 0.000089 & 8.82 $\pm$ 0.24 & 131 \\
2455100.68189 & 23:31:52.5719964 & 0.0000062 & +19:56:13.599894 & 0.000090 & 1.26 $\pm$ 0.04 &  24 \\
2455226.34873 & 23:31:52.5809688 & 0.0000066 & +19:56:13.443962 & 0.000096 & 0.44 $\pm$ 0.04 &  20 \\
2455466.68189 & 23:31:52.6126263 & 0.0000062 & +19:56:13.537918 & 0.000091 & 3.55 $\pm$ 0.16 &  71 \\
2455591.34853 & 23:31:52.6215688 & 0.0000063 & +19:56:13.383583 & 0.000092 & 2.51 $\pm$ 0.14 &  90 \\
2455730.96293 & 23:31:52.6547197 & 0.0000061 & +19:56:13.508659 & 0.000090 & 4.56 $\pm$ 0.13 &  49 \\
2455829.69210 & 23:31:52.6534315 & 0.0000091 & +19:56:13.480797 & 0.000135 & 0.18 $\pm$ 0.03 &  17 \\
2455879.54625 & 23:31:52.6518749 & 0.0000063 & +19:56:13.402746 & 0.000093 & 0.56 $\pm$ 0.03 &  17 \\
2459072.90975 & 23:31:53.0249869 & 0.0000060 & +19:56:12.983043 & 0.000088 & 2.57 $\pm$ 0.04 &  17 \\
2459105.81954 & 23:31:53.0231982 & 0.0000062 & +19:56:12.956925 & 0.000090 & 0.82 $\pm$ 0.03 &  13 \\
2459135.73763 & 23:31:53.0214293 & 0.0000063 & +19:56:12.915578 & 0.000091 & 0.73 $\pm$ 0.04 &  16 \\
\hline
\sidehead{GJ~896B}
\hline
2459105.81954 & 23:31:53.3848609 & 0.0000076 & +19:56:14.482490 & 0.000112 & 0.18 $\pm$ 0.03 &  13 \\
2459135.73763 & 23:31:53.3828962 & 0.0000103 & +19:56:14.447438 & 0.000154 & 0.11 $\pm$ 0.03 &  15 \\
 \enddata
\end{deluxetable}


%
\startlongtable
\begin{deluxetable}{lccc}
\centering                          
\tabletypesize{\scriptsize}
\tablewidth{0pt}
\tablecolumns{4}
\tablecaption{Absolute Astrometric Fits\tablenotemark{a} \label{tab_2}} 
\tablehead{                
\\ Parameters & Single-source  & Single-source$+$acceleration &  Single-companion   
}
\decimals
\startdata
 & (1) & (2) & (3)  \\
\hline                        
& & Fitted Parameters &    \\
\hline                        
$\mu_{\alpha}$ (mas yr$^{-1}$)  &   576.707 $\pm$ 0.014    & 574.171 $\pm$ 0.014  &  574.142 $\pm$ 0.019  \\
$\mu_{\delta}$ (mas yr$^{-1}$)   &  $-$59.973 $\pm$ 0.014  & $-$60.333 $\pm$ 0.014 & $-$60.354 $\pm$ 0.020  \\
$a_{\alpha}$ (mas yr$^{-2}$)      &  $...$   & 0.8893 $\pm$ 0.0034  & 0.8871 $\pm$ 0.0047  \\
$a_{\delta}$ (mas yr$^{-2}$)       &  $...$   & 0.1402 $\pm$ 0.0035 &  0.1400 $\pm$ 0.0049   \\
$\Pi$ (mas)                                 &  161.136 $\pm$ 0.094  & 160.027 $\pm$  0.094 & 159.83 $\pm$  0.13   \\
$P$ (days)          &  $...$  & $...$  & 281.56 $\pm$  1.67 \\
$T_{0}$ (Julian day)    &  $...$  & $...$  &  2,455,696.5 $\pm$ 10.4 \\
$e$                      &  $...$  & $...$  &  0.30 $\pm$ 0.11 \\
$\omega$ (deg)  &  $...$  & $...$  &  344.1 $\pm$ 13.1 \\
$\Omega$ (deg) &  $...$  & $...$  & 47.7 $\pm$ 12.1 \\
$a_{A}$ (mas)           &  $...$  & $...$  & 0.52 $\pm$  0.11 \\
$i$ (deg)             &  $...$  & $...$  & 66.0 $\pm$  15.0 \\
\hline                        
&  & Other Parameters  & \\
\hline                        
$D$ (pc)                                                     &  6.2059 $\pm$ 0.0036   & 6.2489 $\pm$ 0.0037  & 6.2565  $\pm$ 0.0051  \\
$m_{Ab}$ ($M_\odot$)\tablenotemark{b} & $...$  & $...$  & 0.43814 (fixed)  \\
$m_{A}$ ($M_\odot$)                               & $...$  & $...$  & 0.43589 $\pm$ 0.00047  \\
$m_{b}$ ($M_\odot$).                              & $...$  & $...$  & 0.00225 $\pm$  0.00047  \\
$m_{b}$ ($M_{J}$) & $...$  & $...$  &  2.35 $\pm$ 0.49  \\
$a_{Ab}$ (au)                  & $...$  & $...$  & 0.6386 $\pm$ 0.0025   \\
$a_{A}$ (au)           & $...$  & $...$  &  0.00328 $\pm$ 0.00070   \\
$a_{b}$ (au)           & $...$  & $...$  &  0.6353 $\pm$  0.0018 \\
$a_{b}$ (mas)        & $...$  & $...$  &  101.53  $\pm$ 0.42 \\
$\Delta$$\alpha$ (mas)\tablenotemark{c} & 8.51 & 0.24 & 0.14  \\
$\Delta$$\delta$ (mas)\tablenotemark{c}  & 1.51 & 0.24 & 0.13  \\
$\chi^2$, $\chi^2_{red}$\tablenotemark{d}  &   105728.53, 4066.48 &  190.91, 7.95 &  57.43, 3.38   \\
BIC\tablenotemark{e}         & 105,745.86 & 215.17 & 105.93 \\
AIC\tablenotemark{e}         & 105,738.53 & 204.91 & 85.41 \\
%
\enddata
\tablenotetext{a}{The parameters presented here were obtained with AGA. Very similar results were obtained with the least-squares fitting method. Column (1) contain the astrometric fit of the VLBA data without acceleration terms. Column (2) includes acceleration terms. Column (3) corresponds to the full astrometric fit, where a single planetary companion is included.}
\tablenotetext{b}{The mass of GJ~896A, including possible companions, was taken from the combined fit presented in Table 3.}
\tablenotetext{c}{The rms dispersion of the residual.}
\tablenotetext{d}{$\chi^2$ and reduced $\chi^2$ of the astrometric fit. The residuals clearly improve when we include acceleration terms and the companion.}
\tablenotetext{e}{Bayesian Information Criteria (BIC) and Akaike Information Criterion (AIC) are the criteria used to choose the best fitted model \citep[e.g.,][]{liddle07}. The smaller the BIC and AIC values, the better the model. 
}
\end{deluxetable}



\startlongtable
\begin{deluxetable}{lccc}
\centering
\tabletypesize{\scriptsize}
\tablewidth{0pt}
\tablecolumns{4}
\tablecaption{Relative and Combined Astrometry Fits\tablenotemark{a} \label{tab_3}}             
\tablehead{
\\ \colhead{Parameter} &  \colhead{Relative\tablenotemark{b}} & \colhead{Combined\tablenotemark{c}}  & \colhead{Full Combined\tablenotemark{d}}  
}
\startdata
 & (1) & (2) & (3)  \\
 \hline                        
$\mu_{\alpha}$ (mas yr$^{-1}$)  & ... & 571.515 $\pm$ 0.019  & 571.467 $\pm$ 0.023   \\
$\mu_{\delta}$ (mas yr$^{-1}$)  &. ... & $-$37.750 $\pm$ 0.019 &  $-$37.715 $\pm$ 0.023  \\
$\Pi$ (mas)              & 159.88 (fixed) & 159.98 $\pm$  0.14 & 159.88 $\pm$  0.17   \\
\hline                        
& & Binary System &   \\
\hline                        
$P_{AB}$ (days)            &   96606.13  $\pm$ 2.24  & 83665.80  $\pm$ 1.64 & 83664.63  $\pm$ 1.98 \\
$T_{0}$$_{AB}$ (Julian day)\tablenotemark{e}  &  2,504,877.56 $\pm$ 1.48 &  2,401,894.57 $\pm$ 1.00 &  2,401,891.34 $\pm$ 1.19  \\ 
$e_{AB}$                        &  0.0 (fixed)  & 0.108047 $\pm$ 0.000044   & 0.108047 $\pm$ 0.000053  \\
$\omega_{AB}(A)$ (deg)      &    $ ...$        & 127.1531 $\pm$ 0.0037 & 127.1416 $\pm$ 0.0045 \\
$\omega_{AB}(B)$ (deg)      &  0.0 (fixed)  &  307.1531 $\pm$ 0.0037  & 307.1416 $\pm$ 0.0045   \\
$\Omega_{AB}$ (deg)   & 259.7083 $\pm$ 0.0054 & 255.0891 $\pm$ 0.0028 & 255.0919 $\pm$ 0.0034  \\
$a_{AB}(A)$ (mas)            &   ... & 1381.015 $\pm$ 0.069 & 1385.328 $\pm$ 0.083  \\
$a_{AB}(B)$ (mas)          &  ...  & 3676.95 $\pm$ 0.35 & 3672.64 $\pm$ 0.42  \\
$a_{AB}$ (mas)          &  5378.79 $\pm$ 0.51 & 5057.96 $\pm$ 0.36 & 5057.97 $\pm$ 0.43  \\
$i_{AB}$ (deg)            & 130.6592 $\pm$ 0.0087 & 130.0664 $\pm$ 0.0085 & 130.065 $\pm$ 0.010  \\
$Q_{B/A}$                  &  ... &  0.375588 $\pm$ 0.000030 & 0.377202 $\pm$ 0.000037 \\
\hline                        
& & Companion &   \\
\hline                        
$P_{Ab}$ (days)                    & ... & ... &   284.39  $\pm$ 1.47    \\
$T_{0}$$_{Ab}$ (Julian day)             & ... & ... &  2,455,702.65 $\pm$ 17.26 \\
$e_{Ab}$                               & ... & ... &     0.35 $\pm$  0.19  \\
$\omega_{A}$ (deg)            & ... & ... &    353.11  $\pm$ 11.81  \\
$\Omega_{Ab}$ (deg)          & ... & ... &   45.62 $\pm$ 9.60  \\
$a_{A}$ (mas)                    & ... & ... &  0.51 $\pm$ 0.15  \\
$a_{b}$ (mas)            & ... & ... &    102.27   $\pm$ 0.15  \\
$i_{Ab}$ (deg)                      & ... & ... &  69.20 $\pm$ 25.61    \\
\hline                        
&  & Other Parameters &  \\
\hline                        
$D$ ~(pc)                        & ... &  6.2506 $\pm$ 0.0055 & 6.2545 $\pm$ 0.0066  \\
$m_{AB}  ~(M_\odot)$   &  0.544304 $\pm$ 0.000023 & 0.602253 $\pm$ 0.000020 & 0.603410 $\pm$ 0.000025  \\
$m_{AB}(A) ~(M_\odot)$\tablenotemark{f}      & ... &  0.43782  $\pm$ 0.00054 &  0.43814  $\pm$ 0.00065  \\
$m_{AB}(B) ~(M_\odot)$      & ... &  0.16444  $\pm$ 0.00020 & 0.16527  $\pm$ 0.00025 \\
$a_{AB} ~(AU)$         &  33.64265  $\pm$ 0.00052  & 31.615  $\pm$ 0.028  & 31.635  $\pm$ 0.033  \\
$a_{AB}(A) ~(AU)$            & ... &     8.6322   $\pm$ 0.0075     & 8.6646  $\pm$ 0.0091    \\
$a_{AB}(B) ~(AU)$            & ... &   22.983 $\pm$ 0.020   & 22.971  $\pm$ 0.024    \\
$m_{A} ~(M_\odot)$\tablenotemark{g}      & ... & ... &  0.43599 $\pm$ 0.00092  \\
$m_{b} ~(M_\odot)$      & ... & ... &   0.00216 $\pm$ 0.00064 \\
$m_{b} ~(J)$                 & ... & ... &   2.26 $\pm$ 0.57 \\
$a_{Ab} ~(AU)$          & ... & ... &   0.64282  $\pm$ 0.00068  \\
$a_{A} ~(AU)$            & ... & ... &   0.00317 $\pm$ 0.00093    \\
$a_{b} ~(AU)$            & ... & ... &   0.63965  $\pm$ 0.00067    \\
$\Delta$$\alpha$$_{AB}$ ~(mas)\tablenotemark{h} &  123.11  & 89.60  & 89.60   \\
$\Delta$$\delta$$_{AB}$ ~(mas)\tablenotemark{h}  &    89.14 & 74.28 & 74.28   \\
$\Delta$$\alpha$$_{A}$ ~(mas)\tablenotemark{h}   &  ... & 0.27 & 0.21  \\
$\Delta$$\delta$$_{A}$ ~(mas)\tablenotemark{h}    &  ...  & 0.47  & 0.45  \\
$\Delta$$\alpha$$_{B}$ ~(mas)\tablenotemark{h}   &  ... & 0.20 & 0.11  \\
$\Delta$$\delta$$_{B}$ ~(mas)\tablenotemark{h}    &  ...  & 0.12  & 0.042  \\
$\chi^2$, $\chi^2_{red}$\tablenotemark{i}   &  980846.81, 6956.36 & 977729.63, 5785.38 & 977566.99, 6034.36 \\
\enddata
\tablenotetext{a}{The parameters presented here were obtained with AGA. Very similar results were obtained with the least-squares fitting method. The subindex A, B and AB correspond to the main star (GJ~896A), the low mass stellar companion (GJ~896B) and the binary system (GJ~896AB), respectively.}
\tablenotetext{b}{Relative astrometric fit of secondary star GJ~896B around the main star GJ~896A. In this case the parallax is fixed using the solution of the full combined astrometric fit (column 3).}
\tablenotetext{c}{The combined astrometric fit is obtained by fitting simultaneously the relative astrometry of the binary system and the absolute astrometry of the main star GJ~896A and the secondary star GJ~896B (see text). All the free parameters are fitted simultaneously.}
\tablenotetext{d}{The full combined astrometric fit is obtained by fitting simultaneously the relative astrometry of the binary system and the absolute astrometry of the two stars, including a companion associated to GJ~896A (see text). All the free parameters are fitted simultaneously.}
\tablenotetext{e}{Time of the periastron passage. In the case of the relative astrometry (column 1), $T_{0AB}$ corresponds to the time of the passage through the ascending  line of nodes of the orbit of the secondary star around the main star. In the case of the combined astrometry (columns 2 and 3), $T_{0AB}$ corresponds to the time of the periastron passage of the primary star.}
\tablenotetext{f}{Dynamical mass of the star GJ~896A including any posible companions.}
\tablenotetext{g}{Dynamical mass of the star GJ~896A removing the mass of the planetary companion.}
\tablenotetext{h}{RMS dispersion of the residuals. The first two terms correspond to the residuals of the binary system part of the fit, and next two term correspond to the residuals of the absolute astrometry part of the fit of the main star, and the last two terms correspond to the residuals of the secondary star.}
\tablenotetext{i}{$\chi^2$ and reduced $\chi^2$ of the astrometric fit. In all three cases the residuals of the relative astrometry dominates the residuals of the fit.}
\end{deluxetable}

%
\startlongtable
\begin{deluxetable}{lc}
\centering                          
\tabletypesize{\scriptsize}
\tablewidth{0pt}
\tablecolumns{2}
\tablecaption{Mean values and  68$\%$ confidence intervals for the fitted parameters from the $lmfit$ analysis \label{tab_A1}} 
\tablehead{                
\\ Parameters & Fitted Parameters      
}
\decimals
\startdata
$\mu_{\alpha}$ (mas yr$^{-1}$)  &   571.472 $\pm$ 0.018     \\
$\mu_{\delta}$ (mas yr$^{-1}$)   &  $-$37.681 $\pm$ 0.075   \\
$\Pi$ (mas)         &  159.971 $\pm$ 0.036     \\
$P$ (days)         & 83159.87 $\pm$  209.51 \\
$T_{0}$ (Julian day)   &  2,402,034.71 $\pm$ 147.15 \\
$e$                     &  0.1132 $\pm$ 0.0020 \\
$\omega_{A}$ (deg)  &  126.48 $\pm$ 0.11 \\
$\Omega_{A}$ (deg) & 254.855 $\pm$ 0.089 \\
$a_{A}$ (mas)     & 1383.42 $\pm$  5.26 \\
$i$ (deg)             & 129.959 $\pm$  0.015 \\
$Q(m_{B}/m_{A})$            & 0.3773 $\pm$  0.0017 \\
$\chi^2$, $\chi^2_{red}$  &   977668.32, 5785.02   \\
\enddata
\end{deluxetable}

%
%

%

\end{document}